%% file: single.tex
\documentclass[11pt]{article}

\input{full-paper-head}

\usepackage{comment}
\usepackage{etoolbox}
\usepackage{multirow}
\excludecomment{ignore}

\begin{document}

\title{\Large Bounding Cache Miss Costs of Multithreaded Computations \\ Under General Schedulers}
\author{Richard Cole ~\thanks{Computer Science Dept., Courant Institute
of Mathematical Sciences, NYU, New York, NY 10012.
Email: {\tt cole@cs.nyu.edu}.
This work was supported in
 part by NSF Grants CCF-1217989 and CCF-1527568.}
\and Vijaya Ramachandran~\thanks
{Dept. of Computer Sciences, University of Texas, Austin, TX 78712. Email:
 {\tt vlr@cs.utexas.edu}. This work was supported in
 part by NSF Grant CCF-1320675.}
 }

\maketitle

\begin{abstract}
We analyze the caching overhead incurred by a class of multithreaded algorithms when scheduled
by an arbitrary scheduler. We obtain bounds 
that, modulo constant factors,
match or improve 
the well-known $O(Q+S \cdot (M/B))$
caching cost for the randomized work stealing 
(RWS)
scheduler, where $S$ is the number
of steals, $Q$ is the sequential caching cost, and $M$ and $B$ are the cache size and block (or cache line) size
respectively.
\end{abstract}

\input{intro}

\input{table}

\input{results}

\input{theorems}

\input{alg-anal}

\input{ro}

\input{execstack-general}

\input{refined-summary}

\input{general-overview}

\input{bp-general}

\input{hbp-general}

\bibliographystyle{abbrv}
\bibliography{sort,rws-refs}

\appendix
\input{alg-anal-app}

\input{data-examples-app}

\input{local-steal-anal-app}
\input{remote-steal-anal-app}
\input{proof-lem-redistr-app}
\input{pseudo-kernel-app}

\newpage

\end{document}

%% file: full-paper-head.tex
 \usepackage{epsfig}
 \usepackage{amsmath}
 \usepackage{amsfonts}
 \usepackage{amssymb}
 \usepackage{amsthm}

 \usepackage{setspace}
 \usepackage{fullpage}

\usepackage{epsfig}
\usepackage{wrapfig}

\usepackage{enumerate}

\usepackage{soul,xcolor}

 \newcommand{\hide}[1]{}

\newcommand{\Xomit}[1]{}

 \def\qed{\ifmmode$\blacksquare$\else{\unskip\nobreak\hfil
 \penalty50\hskip1em\null\nobreak\hfil$\blacksquare$
 \parfillskip=0pt\finalhyphendemerits=0\endgraf}\fi\vspace{0.3cm}}

 \newtheorem{theorem}{Theorem}[section]
 \newtheorem{lemma}{Lemma}[section]
 
 \newtheorem{definition}[lemma]{Definition}
 \newtheorem{corollary}[lemma]{Corollary}
\newtheorem{fact}[lemma]{Fact}
\newtheorem{example}[lemma]{Example}

 \newcommand{\vone}{\vspace{.1in}}
 \newcommand{\vhalf}{\vspace{.05in}}

 \newcommand{\ceil}[1]{\left\lceil{#1}\right\rceil}

%% file: intro.tex
\section{Introduction}\label{sec:intro}

The design and analysis of 
multithreaded cache-efficient parallel algorithms has been widely studied in recent 
years~\cite{FLPR99,BC+08,CR08,BGS09,CR10,CRSB13}.
Many of these algorithms are based on parallel divide and conquer (called variously 
{\it hierarchical divide and conquer}~\cite{BC+08}, 
{\it hierarchical balanced parallel (HBP) computations}~\cite{CR11,CR12}, etc.).
The performance of these algorithms is usually analyzed for a specific scheduler, especially
with regard to caching costs.

In this paper, we present general bounds on the cache miss cost for several algorithms, when
 scheduled using
 an arbitrary scheduler. Our bounds match the best bounds known for
work stealing schedulers.
The class of algorithms we consider includes efficient multithreaded algorithms
for several fundamental problems
such as matrix multiplication~\cite{FLPR99},
 the Gaussian Elimination Paradigm (GEP)~\cite{CR08},   
longest common subsequence (LCS) and related dynamic programming
problems~\cite{CR08,CLR10}, 
FFT~\cite{FLPR99},
SPMS sorting~\cite{CR10}, list ranking~\cite{CRSB13},  and graph connectivity~\cite{CRSB13}. 
These are all
well-known multithreaded algorithms that use parallel recursive divide and conquer.
Our contribution here is to analyze their
caching performance with
a general scheduler, as a function of the number of parallel tasks
scheduled across the
processors, and to obtain bounds that match the current best bounds 
known only for work stealing schedulers.

We only consider multithreaded algorithms in this paper. As such, we do not directly deal with
related work on parallel, cache-efficient algorithms designed for specific models such as
 the Multi-BSP, Parallel External Memory model,  etc.~\cite{Va08,AGNS08,AGS10,ASZ10,SZ12},
 though all of the algorithms we consider can be scheduled and analyzed on these models.

%% file: table.tex
\begin{table*}[tb]
\begin{center}
\begin{tabular}{|l||c|c|c|}
\hline
\multirow{2}{*}{~~~~~~~~~~{\footnotesize {\bf HBP Algorithm}}} &
{\bf Seq. Cache} & \multicolumn{2}{|c|}{\bf Cache miss bound with $\mathbf{S}$ steals, \boldmath{$C(S) = \min\{A,B\}$}} \\ \cline{3-4}
 & {\bf Bound} {\boldmath $Q$} &  {\footnotesize{\bf  Bound A (Thm. 1)}} & {\footnotesize{\bf Bound B (Thm. 2)}} \\ \hline
\hline
Scan, Prefix Sums  & ${n}/{B}$  & $Q +(M/B) \cdot S$ & $Q + S $   \\ \hline
Matrix Transpose   & ${n}/{B}$  & $Q +(M/B) \cdot S$ & $Q + S \cdot B$   \\ \hline
{$n^3$ Matrix Multiply,  GEP }  & $n^3/(B\sqrt M)$ & $Q +(M/B) \cdot S$ &  $Q + ({n^2}/{B}) \cdot S^{\frac13} +S \cdot B$  \\ \hline
{Strassen Matrix Multiply}    &
 $n^{\lambda}/(BM^{\gamma})$ & $Q +(M/B) \cdot S$  & $Q + ({n^2}/{B})  \cdot S^{1-\frac{2}{\lambda}} + S\cdot B$  \\ \hline
{FFT, ~SPMS, List Ranking} & {$ \frac{n}{B} \cdot \log_M n$} & $Q + (M/B) \cdot S$ & $Q +  \frac{n}{B} \cdot \frac{\log n}{\log [(n \log n)/S]} + S\cdot B$ \\ \hline
Graph Connectivity & $\frac{n+m}{B} \cdot \log_M n$ & $Q + (M/B) \cdot S$ &
     $Q +  \frac{n+m}{B} \cdot \frac{\log^2 n}{\log [((n+m) \log n)/S]}+ S\cdot B$ \\ \hline
Finding LCS sequence & $n^2/(BM)$  & $Q + (M/B) \cdot S$
   & $Q + ({n}/{B}) \cdot  \sqrt S + S\cdot B$ \\ \hline
     
\end{tabular}
\end{center}
\caption{Our upper bound for cache miss cost,
$C(S)$,  with $S$ steals, for a general scheduler;  $O(\cdot)$ is omitted.
The sequential cache miss bound is $Q$, and a tall cache is assumed.
Always, the new bound for a general scheduler matches or improves the bound 
in~\cite{Blumofe96b,FS06},
modulo a constant factor,
and matches the
bound in \cite{CR12b}; all of these prior bounds held
only for work stealing. 
For Strassen, $\lambda=\log_2 7$ and $\gamma = (\lambda/2) -1$~\cite{FLPR99,CR11}.
}
\label{table1}
\end{table*}

%% file: results.tex
  \subsection{Related Work}\label{sec:related}

  Let $Q$ be the sequential caching cost of 
a multithreaded computation,  and
  let $C(S)$ be the 
  caching cost incurred in a parallel execution with $S$ steals. 
Blumofe et al.~\cite{Blumofe96b}
observed that an execution of a
 computation that incurs $S$ steals when scheduled
under randomized work-stealing (RWS) can be partitioned into $O(S)$ fragments,
where each fragment runs on a single processor in this parallel execution, and represents a
 contiguous portion of the sequential execution of the computation. They  then observed 
 that the computation regains the state of the sequential execution after reading at most $M/B$ distinct
 blocks, and thereafter inherits the sequential cache complexity. Thus, 
 $C(S)$ is bounded by
$Q + O(S \cdot M/B)$,
 where $Q$ is the cache miss bound for
 a sequential execution.
Although not explicitly stated, this observation
appears to need the assumption
that all variables have a fixed allocation, regardless of the amount of parallelism.
In~\cite{CR10}, we showed how to account for dynamically allocated variables
(the issue being that
 block boundaries on the execution stack for a stolen task can diverge from those at the parent task),
giving a bound of $O(Q + S \cdot M/B)$ on $C(S)$.
 
  Frigo and Strumpen~\cite{FS06} considered the above set-up 
for computations where any  fragment of
  size $r$ that occurs in a parallel execution incurs $O(f(r))$ cache misses, for some concave function
  $f$. They then showed that some of the known cache-efficient multithreaded algorithms have good concave
  functions $f$ satisfying the above property and used this to refine the bound 
  in~\cite{Blumofe96b}.
    If
  a multithreaded algorithm makes calls to different subroutines with different cache complexities, then
  the concave function will be at most as good as the least efficient of the caching bounds. Thus,
  the results in~\cite{FS06} are most effective for cache-efficient algorithms that recursively 
  call only themselves, such as the matrix multiplication algorithm with depth $n$ (Depth-$n$-MM), the 
  Gaussian Elimination
  paradigm (IGEP)~\cite{CR08}, 
  and stencil computations. Further, even for these algorithms, the results in~\cite{FS06}
  apply only if parent stealing is used (i.e., if the node forking the two parallel tasks is placed on the task
  queue, as is the case in Intel CilkPlus). In~\cite{CR12b} an example is given where the result in~\cite{FS06}
  for Depth-$n$-MM  does not hold under child stealing (where the right child of the forking
  node is placed on the task queue, as in Intel TBB and Microsoft PPL).
  
  The bounds in~\cite{FS06} were matched and also extended to a more general class of
 HBP computations 
 for RWS under child stealing in~\cite{CR12b}. 
   The methodology in~\cite{CR12b}
    is to charge the cost of the cache miss 
overhead to $O(S)$ disjoint tasks in the sequential computation,
    where each task is an HBP sub-computation, 
 and then to bound the cost of the worst-case configuration of such a collection of $O(S)$ disjoint tasks.

    These prior results were reported only for RWS, but the analysis holds for any work stealing scheduler. 
Work stealing is a natural and effective method
    for scheduling multithreaded algorithms, and is implemented in
CilkPlus, TBB and PPL, as noted above.
A key feature of work-stealing is that the 
task that an idle processor steals (i.e., moves) from another processor is the one at the head of 
the other processor's task queue. In other words, tasks are stolen from 
the task queue at any given processor in FIFO order.
However, a multithreaded   algorithm may  be scheduled in
environments where a  work stealing scheduler is not available. 
    In such a case,
the system scheduler will be used to schedule the parallel tasks
and this
 scheduler may not necessarily schedule tasks in FIFO order. 
For instance, SJF (Shortest Job First) is a commonly used scheduling policy, and this policy need not be FIFO at each processor. The Linux scheduler uses the Completely Fair Scheduler, and it is not clear if that 
scheduler
uses the FIFO needed for work stealing.
    
Another reason for considering a general scheduler is 
to obtain `oblivious' results as in 
 sequential cache-oblivious algorithms~\cite{FLPR99}, network-oblivious algorithms for distributed memory~\cite{BPPSS16}, 
and multicore-oblivious \cite{CRSB13} and 
 resource-oblivious \cite{CR10,CR12} algorithms for shared memory multicores. In all of these cases the desire is to have algorithms analyzed in a machine-independent manner so that bounds hold across diverse platforms. In that spirit our results give scheduler-independent results that extend across all types of schedulers as long as there is no preemption, duplication of tasks, or failures.

Further, one could consider future scenarios where new criteria such as power consumption may dictate the need for new types of schedulers. Our results show that there is not much degradation in the caching performance as a function of the number of parallel tasks scheduled even if such schedulers do not steal from the top of the dequeue.

If the scheduler
      does not  steal in FIFO order, then the analysis used to derive  the earlier results
    for caching overhead when using
 RWS is not valid.
 Thus,
new techniques need to be developed in order to analyze caching costs
with a general scheduler.
This is the topic of this paper.

In this paper, we show that for a general class of multithreaded algorithms, including all 
those
with series-parallel fork-join calls,
 the cache miss excess 
remains bounded by
$O(Q + S\cdot M/B)$, and that
for a class of well-structured HBP algorithms 
(including
    those listed in Table~\ref{table1}), the cache miss excess is
 bounded by the best bound currently known
 for work stealing schedulers.  
 
  We are able to achieve good bounds even
  when considering the worst-case effects of `false-sharing' (fs misses)
  as long as we use
the algorithms with the small modifications given in~\cite{CR12};
we omit discussing this 
here.

\subsection{Overview of Our Results}\label{sec:results}
 
We assume a tall cache ($M \geq B^2$), and we assume that
a sequential execution that accesses $r$ data items  accesses 
  $O((r/B) + \sqrt r)$ blocks (see Sections~\ref{sec:bp}, \ref{sec:hbp-outline}).
Our main results are Theorems~\ref{thm:cache-miss-simple} and
  \ref{thm:cache-miss-overhead-gen} in Section~\ref{sec:main-thms}, and 
  Table~\ref{table1} lists the bounds we obtain
   for some well-known algorithms
  by applying these two theorems.
   All of these algorithms are well-known
parallel multithreaded algorithms,
    and all have excellent sequential cache-oblivious
   caching bounds. 
   
Consider a computation whose parallel execution incurs $S$ steals.
Previous analyses 
for the cache miss overhead
all took the following approach: the sequential execution was partitioned
into 
$O(S)$
consecutive pieces 
or fragments, which we call \emph{task kernels},
with the property that in the parallel
execution each task kernel was executed on a single processor.
 Then the analyses amounted
to bounding the 
amount of data a task kernel uses that was used by an earlier task kernel in the sequential execution
and which could have been available in the cache;
this upper bounds the additional reloads due to steals.
However, with a general scheduler, a partitioning with these properties in not possible in general,
as we show
 in Example~\ref{ex:gen-sched-interleave} in Section~\ref{sec:cache-overhead}.
Nonetheless, we are able to recover the simple
$O(Q + S \cdot M/B)$ bound on 
$C(S)$,
the number of cache misses
with $S$ steals.
Further, with a more sophisticated analysis we achieve the 
results in Bound B
in Table~\ref{table1},
bounds 
that
match the 
earlier
results in~\cite{CR12b} which hold only for RWS, and which
can be a strict improvement (depending on the value of $S$)
over the $O(Q + S \cdot M/B)$ bound,
 as shown
in Section~\ref{sec:alg-anal} for FFT and SPMS sorting.
   
   At a high level, our approach to establishing our bounds is similar to the one used in~\cite{CR12b}
   for work stealing schedulers. It bounds the caching overhead for an HBP computation incurring
   $S$ steals as being no more than the cost of reloading the cache for $O(S)$ HBP tasks
   in the computation. The final bound is obtained by considering the worst-case cost
   for a collection of $O(S)$ HBP tasks in the computation. However, within this high level approach, 
   our current method differs from the one in~\cite{CR12b}, as described below.
   
    In~\cite{CR12b}, the $O(S)$ 
tasks
  were required to be disjoint tasks
(as was the case in~\cite{Blumofe96b,FS06} as well),
and this resulted in several different case analyses for 
   different types of HBP computations.
It  also required rather strong balance conditions for the
   sizes of sibling recursive tasks
because costs were being allocated from a steal-incurring subtask to a steal-free sibling. 
In our current analysis, we allow these $O(S)$ 
distinct HBP tasks to overlap, and we allocate the costs to the steal-incurring
task itself.
This allows us to unify the analysis for all HBP computations  into a single argument.

   \vone
   \noindent
\emph{Organization of the Paper.} In Section~\ref{sec:main-thms} we 
 state our two main theorems, and we describe the concrete results we obtain from our second theorem for specific algorithms.
 Section~\ref{sec:sched} gives basic background on work stealing and scheduling parallel tasks, and Section~\ref{execstack-general}
 describes our set-up for general schedulers. In Section~\ref{sec:cache-summary} we define {\it task kernels} and give a proof of our
 first main theorem
 (Theorem~\ref{thm:cache-miss-simple}). Finally, in Section~\ref{sec:general} we present our refined analysis for BP and HBP computations,
 and establish our second main theorem.  
 Some of the details and proofs are deferred to
 the appendix.

%% file: theorems.tex
\section{Our  Main Theorems}
\label{sec:steps2-3}
\label{sec:main-thms}

We consider a shared memory parallel
 environment comprising $p$ processors,
 each with
 a private cache of size $M$. The $p$ processors communicate through an arbitrarily large shared memory.
 Data is organized in blocks (or `cache lines') of size $B$.
 
We will express parallelism through paired fork and join operations.
A fork spawns two tasks that can execute in parallel. Its corresponding join is a
synchronization point: both of the spawned tasks must complete before the computation
can proceed beyond this join.
For an overview of this model, see Chapter 27 in~\cite{CLRS09}.

Our first theorem applies to the cache miss overhead under the
scheduling of any series-parallel computation dag by a general
scheduler, and it generalizes an earlier result
in~\cite{Blumofe96b} that was established for Cilk (i.e., for RWS).

\begin{theorem}\label{thm:cache-miss-simple}
Let $\cal A$ be 
a series-parallel
 algorithm and suppose it incurs $S$ steals
in a parallel execution using a general scheduler.
Then the cache miss cost of this execution is 
$C(S) = 2Q + O(S \cdot M/B)$,
where $Q$ is the number of cache misses incurred by $\cal A$ in a sequential execution.
\end{theorem}

Our second theorem improves on the above theorem
for the following
class of algorithms,
based on~\cite{CR11,CR12b}. 
Here, given a task $\tau$, its {\it size}, $|\tau|$,  is 
 the number of distinct data items read or written by $\tau$; 
this excludes any local variables declared by $\tau$. A
{\it balanced fork-join computation} consists of a fork tree followed by
a join tree on a common set of leaves, where the sizes of the tasks 
decrease geometrically from parent to child in the fork tree.

\begin{definition}
\label{def:hbp}
{\bf (HBP task)}
A BP  algorithm
(or task)
is a balanced  binary fork-join computation on $n$ leaves, where each fork, join and leaf node performs
$O(1)$ computation.

A Type 1 
HBP 
task comprises a sequence of $O(1)$
BP tasks.

A Type $k$
HBP task,
 for $k \geq 2$,
 comprises a sequence of 
$O(1)$
 \emph{constituent} tasks.
Each constituent task is either a BP task, a
Type $h<k$ HBP task, or 
a {\it recursive constituent}, which is
an ordered collection of 
one or more
recursive instances of the 
Type $k$ task.
Each such ordered collection is initiated by a
binary fork tree and ended by a complementary join tree.

In addition, certain requirements apply to data layout and data accesses as described in Section~\ref{sec:data}.
\end{definition}

In order to bound the additional cache misses incurred due to steals, we now define $x(\tau)$,
the \emph{extended size} of $\tau$,
as follows.

\begin{definition}
\label{def:extended-size}
{\bf (Extended size)}
Let $\tau$ be a task that calls $\tau_1, \tau_2, \ldots, \tau_l$,
where each  $\tau_i$ is
a BP constituent task or an individual recursive task forked by a 
recursive constituent of $\tau$.
Then, $\tau$'s extended size
$x(\tau)$
is given by
$x(\tau) = |\tau| + \sum_{i=1}^l |\tau_i|$,
\end{definition}
The extended size of a task $\tau$ incorporates $\tau$'s size, $|\tau|$, together with 
the sizes of individual tasks in its constituent tasks.
The additional term
over $|\tau|$ is the sum of the
sizes of the tasks called by $\tau$.  
This is done in order to account for the fact that a stolen sub-task of $\tau$ may need to read again
some of this data, and $\tau$ can have several stolen sub-tasks. 
Also, there may be overlap in the data 
accessed by different tasks called recursively by $\tau$.
 In general, in the extended size of $\tau$ the individual sizes of the
 tasks called by $\tau$
are added to the size of 
$\tau$, possibly resulting in a value much larger than $|\tau|$.
However, for 
all the algorithms we consider (see Table~\ref{table1}),
 the value of $x(\tau)$ remains $O(|\tau|)$.

We now
 state the constraints 
we impose
 on the algorithms we consider.
To achieve the strongest cache miss bounds we need the algorithm to be cache-compliant, as defined next.

\begin{definition}\label{def:good-hbp}
{\bf (Cache-compliant task)}
An HBP 
task $\cal A$ is 
\emph{cache-compliant} if for each recursive task $\tau$ in $\cal A$ and for each recursive call $\tau'$
made by $\tau$, there is a constant $\alpha <1$ such that
\begin{itemize}
\item
$|\tau'| \le \alpha |\tau|$,
\item
$x(\tau') \le \alpha \cdot x(\tau)$, and
\item
$\tau$ makes $O(|\tau|)$ recursive calls.
\end{itemize}
\end{definition}

All the algorithms we consider  are cache-compliant.

Our second theorem,
given below, provides a 
refined bound for $C(S)$ for HBP algorithms 
(Table~\ref{table1} gives a
 tighter result for Scan and Prefix Sums 
that
does not follow directly from Theorem~\ref{thm:cache-miss-overhead-gen};
that result
is shown
 in Section~\ref{sec:bp}).

\begin{theorem}
\label{thm:cache-miss-overhead-gen}
Suppose the execution of 
a cache-compliant
Type $k$
HBP algorithm $\cal A$ incurs $S$ steals when executed using a general
scheduler. 
Suppose that in a sequential execution $\cal A$ incurs $Q$ cache misses.\\
(i) If $k=1$ then $C(S) = O(Q + S \cdot B)$.\\
(ii) If $k \geq 2$ then
 there is a collection $\tau_1, \tau_2, \ldots, \tau_l$ of distinct recursive tasks,
with $l = O(S)$,
where each of the $\tau_i$ is an $h$-HBP task for some $ 2 \le h \le k$, including possibly the whole computation,
such that the cost, $C(S)$,  of the cache misses
 incurred by this execution of $\cal A$ is bounded by
\[
C(S) = O\left(Q + (\sum_{i=1}^l x(\tau_i)/B )+ S\cdot B\right).
\]
\end{theorem}

With the above bound in hand, it will suffice to bound 
$\sum_i x(\tau_i)/B$, 
where the sum is
over all 
$l$
 tasks
specified in Theorem~\ref{thm:cache-miss-overhead-gen}.
The result is a bound on 
$C(S)$ that is never worse than the earlier bound of
$O(Q + S \cdot M/B)$,
and in some cases improves on it,
and which applies not only to work stealing schedulers but
also to general schedulers.

As in~\cite{BL99,ABB02}, we
 can incorporate the above bound for 
 the overall cache miss cost $C(S)$ for
 any scheduler that steals $S$ tasks into a bound on the
overall time for the parallel execution as follows. Let $b$ be the cost of a cache miss, and
$s$ the cost
of a steal, i.e., $s$ is the time taken by the scheduler to transfer a parallel task from its original processor
to another processor that will execute it in parallel.
Let $T_1$ be the sequential execution time for the computation,
let $T_{\infty}$ be the  span (or critical path length)
  of the parallel computation,
and let $I$ be the total time spent by  processors idling while not computing, 
stealing, or waiting on a cache miss.
Then the time taken by this parallel execution is
given by:

\begin{equation*}
\label{eqn:time}
T_p = \frac{1}{p} \left( T_1 + b \cdot  C(S) + s \cdot S + I \right)  + b \cdot T_{\infty}.
\end{equation*}

In the above equation, $S$ and $I$ depend on the scheduler: A well-designed scheduler would
steal as few tasks as it can while keeping all processors engaged in computation. Our
contribution in this paper is to obtain a good bound for cache-compliant HBP
computations for the term $C(S)$ in the above equation.

At a  high level, 
our analysis proceeds as follows.
It identifies $O(S)$ `special' recursive tasks, some of which may be nested one in another,
and assigns to these special tasks
all the cache-miss costs apart from the sequential execution cost.
In addition, each steal will be assigned to a special task (this task will
be said to \emph{own} the steal).
Let $\tau_i$ be one of these special tasks and suppose it owns $S_i$ steals;
then the costs assigned to $\tau_i$ will be bounded by 
$O(x(\tau_i)/B + S_i \cdot B)$
as we will see later.

%% file: alg-anal.tex
\subsection{Analysis of Specific Algorithms}
\label{sec:alg-anal}
We apply Theorem~\ref{thm:cache-miss-overhead-gen} to several well-known algorithms,
to obtain the results for bound B in Table~\ref{table1}. The GEP and LCS algorithms are presented in~\cite{CR07,CR10}, while the others
 are described in~\cite{CR12}
(where their false sharing costs
are analyzed).
Here we obtain bound B for a couple of entries listed in Table~\ref{table1}. For the remaining entries in the table 
see Section~\ref{sec:alg-anal-app} in the appendix.

\vhalf
\noindent
{\bf $\log^2 n$-MM.} This is a
Type 2
HBP that has one recursive constituent that makes 8 recursive calls to $n/2\times n/2$ matrices,
 and a BP task that adds up the outputs of the recursive calls in pairs. Its sequential cache complexity is $O(n^3 / (B \sqrt M)$.
 Applying 
Theorem~\ref{thm:cache-miss-overhead-gen} we see that 
 the $l$ largest HBP tasks are obtained
by including all recursive tasks up to $j=(1/3) \log l$ levels of recursion. The sum of the sizes of these tasks is
$O((n^2/4^j) \cdot 8^{(1/3) \log l})
 = O(n^2 \cdot l^{1/3})$. Since $l=O(S)$,  we obtain
  the overall cache miss cost with $S$ steals as
 $O(Q + (n^2/B) \cdot S^{1/3} + S \cdot B)$.
 
\vhalf
\noindent
{\bf FFT, SPMS Sort.} The algorithms for both FFT~\cite{FLPR99} and 
SPMS sort~\cite{CR10} have the same structure, being Type 2 HBP algorithms that 
recursively call two collections of $O(\sqrt n)$ parallel tasks of size $O(\sqrt n)$, together
with a constant number of calls to BP computations.

To bound
$ \sum_{i=1}^{O(S)} |\nu_i|/B$, 
we observe that the
total size of tasks of size $r$ or larger is $O(n \log_r n)$, and there are
$\Theta(\frac{n}{r} \log_r n)$ such tasks.
Choosing $r$ so that $S = \Theta(\frac{n}{r} \log_r n)$
we obtain $r\log r = \Theta( n\log n/S)$, so
$\log r = \Theta(\log( [n \log n/S])$.
Thus  $ \max_{\cal C} \sum_{\nu_i\in {\cal C}}\frac{|\nu_i|}{B}
= O(\frac{n}{B} \log_r n) = O(\frac{n}{B} \frac{\log n}{\log [(n\log n)/S]})$.

The analysis for SPMS is very similar, except that we need to handle two BP computations
with somewhat irregular access patterns. 

\vhalf
\noindent
{\bf Observation.}
For FFT and SPMS, our refined bound is strictly better than the 
$O(Q + S \cdot M/B)$
 bound since
our overhead remains $O(Q)$ when $M ^{\epsilon} = O((n \log n)/S$ for any constant $\epsilon>0$, 
while the simple bound needs $M \log M = O((n \log n)/S)$ for $Q$ to dominate $S \cdot M/B$.

%% file: ro.tex
 \section{Scheduling Parallel Tasks}\label{sec:sched}

The {\it computation dag} for a computation on a given input is
the acyclic graph that results when we have a vertex (or node)
for each unit (or constant) time computation, and
a directed edge from a vertex $u$ to a vertex $v$ if vertex $v$ can begin its computation 
immediately after $u$ and $v$'s other predecessors complete their computations,
but not before. Since we consider multithreaded algorithms with binary forking, where the fork-joins are nested, the computation dag is a series-parallel graph.

During the execution of a computation dag  on a given input, a parallel task is created each time a
fork step $f$  is executed.
At this point the main computation proceeds with the left child of 
$f$, as in a standard sequential dfs computation,
while  the task $\tau$ at the right child $r$
of the fork node is made available
to be scheduled in parallel with the main computation. 
(This is  
{\it child stealing}, or
the {\it help-first policy}~\cite{GZ+09}; one could also use the
 work-first policy where the main computation proceeds with the 
task spawned at the right child.)
The parallel task $\tau$ consists of all of the computation
starting at $r$ and ending at the step before the join corresponding to the fork step $f$.
A run-time scheduler determines if a forked task is to be moved to 
another processor for execution in parallel.

\subsection{Caching Overhead Under Work-stealing}\label{sec:cache-cost}

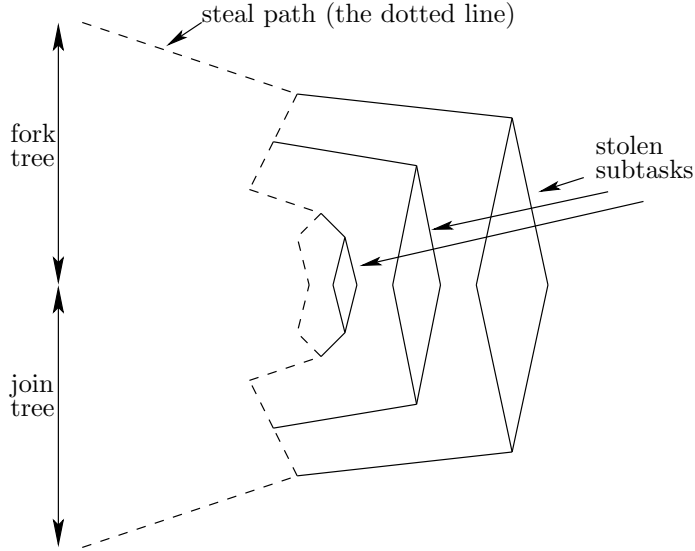
\begin{figure}[htbp]
\begin{center}

\begin{picture}(0,0)%
\includegraphics{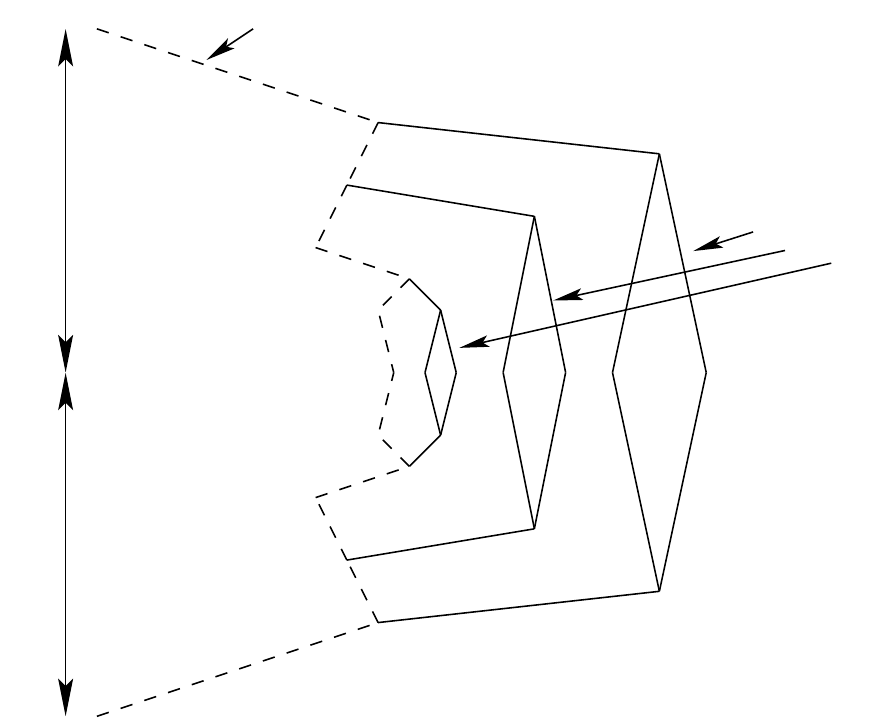}%
\end{picture}%
\setlength{\unitlength}{3947sp}%
\begingroup\makeatletter\ifx\SetFigFont\undefined%
\gdef\SetFigFont#1#2#3#4#5{%
  \reset@font\fontsize{#1}{#2pt}%
  \fontfamily{#3}\fontseries{#4}\fontshape{#5}%
  \selectfont}%
\fi\endgroup%
\begin{picture}(4243,3447)(211,-2773)
\put(3901,-286){\makebox(0,0)[lb]{\smash{{\SetFigFont{10}{12.0}{\rmdefault}{\mddefault}{\updefault}{\color[rgb]{0,0,0}stolen}%
}}}}
\put(1426,539){\makebox(0,0)[lb]{\smash{{\SetFigFont{10}{12.0}{\rmdefault}{\mddefault}{\updefault}{\color[rgb]{0,0,0}steal path (the dotted line)}%
}}}}
\put(3901,-436){\makebox(0,0)[lb]{\smash{{\SetFigFont{10}{12.0}{\rmdefault}{\mddefault}{\updefault}{\color[rgb]{0,0,0}subtasks}%
}}}}
\put(226,-211){\makebox(0,0)[lb]{\smash{{\SetFigFont{10}{12.0}{\rmdefault}{\mddefault}{\updefault}{\color[rgb]{0,0,0}fork}%
}}}}
\put(226,-361){\makebox(0,0)[lb]{\smash{{\SetFigFont{10}{12.0}{\rmdefault}{\mddefault}{\updefault}{\color[rgb]{0,0,0}tree}%
}}}}
\put(226,-1786){\makebox(0,0)[lb]{\smash{{\SetFigFont{10}{12.0}{\rmdefault}{\mddefault}{\updefault}{\color[rgb]{0,0,0}join}%
}}}}
\put(226,-1936){\makebox(0,0)[lb]{\smash{{\SetFigFont{10}{12.0}{\rmdefault}{\mddefault}{\updefault}{\color[rgb]{0,0,0}tree}%
}}}}
\end{picture}%

\caption{\label{fig:steal-path}A steal path for a work-stealing scheduler
with three stolen subtasks of a BP task.
}

\end{center}
\end{figure}

An important class  of schedulers is the work-stealing scheduler. This is a distributed scheduler
for multithreaded computations.
Each processor maintains a task queue on which it enqueues the parallel tasks it generates.
When a processor is idle it attempts to 
steal, i.e.,
obtain a parallel task from
the head of the task queue of another processor.
The exact method for identifying
the processor from which to obtain
an available parallel task determines
the type of work-stealing scheduler being used; however the stolen task is always the task at
the head of the task queue of the chosen processor.
The most popular type is randomized work-stealing (RWS, see e.g.,~\cite{BL99}), where a
processor picks a random processor and steals
the task at the head of its task queue, if there is one.
Otherwise, it continues to pick random processors and tries to find an available parallel task
until it succeeds, or the computation completes. RWS has been widely analyzed and used,
notably in Cilk.
The following  is a well-known fact about work stealing schedulers
(see Figure~\ref{fig:steal-path}).

\begin{fact}\label{fact:stealpath}
{\bf The Steal Path for Work-stealing Schedulers).}
Let $\tau$ be either the original task or a stolen subtask.
Suppose that $\tau$
incurs steals of subtasks $\tau_1, \cdots , \tau_k$.
Then there exists a path $P_{\tau}$ in $\tau$'s computation dag from its
root to its final node such that the parent of every stolen
task $\tau_i$ lies on $P_{\tau}$, and every off-path right child of
a fork node on
$P$ is the start node for a stolen subtask.
\end{fact}

\begin{ignore}
\noindent
{\it Comment.}
The above fact follows by observing the nature of a depth-first-search
based sequential execution of recursive computations.
\end{ignore}

We now review a well-known bound for RWS
in Blumofe et al.~\cite{Blumofe96b}
on $R(S)$, the caching overhead
(over and above the sequential caching cost) in a parallel execution that incurs $S$ steals.
We will assume an optimal offline
cache replacement policy 
(LRU was assumed in~\cite{Blumofe96b} but that result clearly also applies to an optimal policy\footnote{Acar et al.~\cite{ABB02} later reproved this result for simple caches.}).
Let $\sigma$ be the sequence of steps executed in a sequential execution.
Now consider a parallel execution that
incurs $S$ steals.
Partition $\sigma$ into contiguous portions so that in the parallel execution
each portion is executed in its sequential order on a single processor.
Then, each processor can regain the state of the
cache in the sequential execution once it has
accessed $M/B$ distinct blocks during its execution.
Thus if there are $K$ portions, then 
there will be at most $R(S) = K \cdot M/B$ additional cache misses.
(Actually, if there are dynamically allocated variables, the complete
justification of this claim requires the parallel analogue of
the regularity assumption formulated in~\cite{FLPR99},
since block boundaries may change across processors;
see~\cite{CR10}.)

It is shown 
in~\cite{Blumofe96b} that $K=2S$ for Cilk, i.e., for a work-stealing scheduler.
This is readily seen from 
Fact~\ref{fact:stealpath}.
 A steal
creates three fragments within the sequential computation --- (1)  the sequential computation up to 
when the stolen task would start its computation, (2) the computation of the stolen task, which will
occur on a different processor, and (3) the sequential computation following the stolen task. Each
of these three fragments is computed sequentially. Each successive steal creates two additional
fragments leading to
$K=2S$
sequential fragments 
beyond the initial fragment up to the first steal,
and hence
$R(S) \leq 2S \cdot M/B$
additional cache misses.
This implies
 $C(S) \leq Q + 2S \cdot (M/B)$ under a work-stealing scheduler.

%% file: execstack-general.tex
\section{General Schedulers}\label{execstack-general}

In this paper, 
we consider the cache miss overhead for
  a general scheduler that is not necessarily work stealing. We will assume that there is
no redundancy in the computation, and 
that
each node in the computation dag is executed at
exactly one processor. We will view the general scheduler as being similar to a work stealing scheduler, except that the task stolen from the chosen processor can be an arbitrary
parallel
task available for computation, not necessarily the task at the head of its task queue.

When a task other than the topmost task on a task queue is stolen, we
call this a \emph{deep} steal.
More precisely, we have the following definition. 

\begin{definition}\label{def:deep-steal}
{\bf (Deep steal)}
Let $\sigma$ be a task stolen from the task queue, $\Pi$, of $\tau$, and suppose that $\sigma$ is not the
first task placed on $\Pi$. Let $\sigma'$ be the task placed on $\Pi$ immediately before $\sigma$.
Then, this steal of $\sigma$ is a \emph{deep steal} if $\sigma'$ is not stolen from $\Pi$.
\end{definition}

In order to essentially maintain 
Fact~\ref{fact:stealpath} (the Steal Path Fact)
we will treat all the tasks ahead
of $\sigma$ on the task queue $\Pi$
as if they were
`pseudo-stolen' as per
 the following definition.

\begin{definition}\label{def:pseudo}
{\bf (Pseudo-stolen task)}
Consider the task queue $\Pi$ for a task $\tau$, and let $\sigma$ be stolen from $\Pi$ as a deep steal. Any task that was placed on $\Pi$
before $\sigma$ and  which
remains unstolen is a \emph{pseudo-stolen} task.
\end{definition}

We observe that in a computation that incurs deep steals, the steal path 
 will contain not only the parent of
every stolen task but also the parent of every pseudo-stolen task.

\subsection{Execution Stacks}
\label{sec:exec-stack}

In order to obtain a tighter 
bound on the additional 
costs due to steals,
 we 
now
  take a closer look at how one stores variables that are generated during the execution of the
algorithm.
It is natural for the original task and each stolen task to each have an
execution stack on which they store the variables declared by their residual task.
This collection of stacks in a Cilk implementation is referred to as a distributed {\it cactus stack}
in~\cite{Blumofe96b}.
$E_{\tau}$ will denote the execution stack for a task $\tau$ if it has one.
As is standard, each procedure and each fork node stores the variables
it declares in a segment
on the current top of the execution stack for the
task to which it belongs,
following the usual mode for a procedural language.

\vhalf
\noindent
{\bf Execution Stack and Task Queue.}
The parallel tasks for a processor $P$ are enqueued on its task queue in the order in which 
the segments for their parent fork nodes
 are created on $P$'s execution stack. The task queue is a double-end queue,
and $P$ will remove an enqueued task $\sigma$ from
 its task queue when it begins computing on
$\sigma$'s segment.

As noted in Section~\ref{sec:sched},
work stealing is a popular scheduling strategy where a task that is stolen (i.e.,
transferred to another processor) is always the one that is at the
head of the task queue in the
processor from which it is stolen.
However, with a general scheduler, an arbitrary task on the task queue can be stolen.

\vhalf

\noindent
{\bf Cache Misses when Accessing an Execution Stack}
Suppose that  a subtask $\tau'$ is stolen from a task $\tau$.
Consider the join node $v$
immediately following the node
at which the computation of $\tau'$ terminates.
Let $P$ be the processor executing $\tau - \tau'$ when it reaches node $v$ and
let $P'$ be the processor executing $\tau'$ at this point.
To avoid unnecessary waiting, whichever processor (of $P$ and $P'$) reaches $v$ second
is the one that continues executing the remainder of $\tau$.
If this processor is $P'$, we say that $P'$ has usurped
 the computation of $\tau$.
The effect, in terms of cache misses, is that in order to access variables
on $E_{\tau}$, $P'$ will incur cache misses that $P$ might not.
Even if $P'$ does not usurp $\tau$, 
$P$ may have to read additional data
when continuing the execution of $\tau$ beyond $\tau'$ 
(due to its having been first read by the stolen subtask).
Our analysis of cache miss overhead in a parallel execution
 will use a single method to cover the costs in both cases.
This analysis assumes that no data is in cache at the start of the execution of
$\tau$ beyond $\tau'$, whether $P$ or $P'$ is performing this execution, and hence can
 only overestimate the necessary reloads of data.

\vhalf
\noindent
{\bf Execution Stacks for a General Scheduler.}
We observe that when using a general scheduler,
additional execution stacks may be needed.
For suppose that processor $P$ is executing a task $\tau$ from which a deep steal of subtask
$\sigma$ occurs.
Let $P'$ be the processor executing $\sigma$.
Suppose that $P$ is the first of $P$ and $P'$ to reach the join node at which the steal ends.
Then $P$ will leave the continuation of the execution of $\tau$ to  $P'$.
But $P$ needs to continue the execution of the
parallel tasks still on its task queue,
performing them in dfs order (i.e.\ from the 
rear of the queue).
However, $P$ cannot use the execution stack for $\tau$ to store the variables
for these subtasks as (a) this would violate the standard practice in which the current variable
order on the execution stack corresponds to the current path of open procedure calls,
and (b) additional space on this stack may be needed for the execution by $P'$ of the portion of
$\tau$ following the join node.
In these circumstances $P$ will create a new execution stack for each such
pseudo-stolen task on its 
task queue as and when it starts its execution.
This is exactly what would happen were the task to be stolen.
The difference is that this will not count as a steal.
In fact, whether $P$ reaches the join node first or not, it will need to execute the
tasks remaining on its task queue, and it will use a new execution stack for each
such pseudo-stolen task.
$P$ will continue to have a single task queue, however.
The collection of execution stacks can still be viewed as a cactus stack~\cite{Blumofe96b},
with one branching for each stolen or pseudo-stolen task.

\subsection{Caching Overhead for General Schedulers}
\label{sec:cache-overhead}

We now give an example of an execution where
a task could be fragmented into a sequence of several non-contiguous fragments of
execution 
due to a single steal,
and hence 
the analysis  
in~\cite{Blumofe96b}
for the cache miss excess bound of  $R(S)=O(S \cdot M/B)$,
which we saw earlier for work-stealing schedulers,
 does not immediately hold.

\begin{example}
\label{ex:gen-sched-interleave}
\normalfont{\rm See Figure~\ref{fig:pseudo-stolen}.
Let $\tau$ be a 
balanced
fork-join task with $n$ leaves,
with unit-cost computation at each node.
Suppose 
$\tau$
incurs one steal of a subtask $\tau_1$,
where the start of $\tau_1$ 
is reached by traversing a path $\cal P$ of
$k = (\frac 12 \log n) - 1$ left child links followed by one right child link.
Let $\mu_1, \mu_2, \ldots, \mu_k$ be the right subtasks of path $\cal P$,
from top to bottom, preceding $\tau_1$.
Note that 
each of the tasks $\mu_1, \mu_2, \ldots, \mu_k$ is a pseudo-stolen task.
Let $\overline{\cal P}$ be the path in the join tree
complementary to $\cal P$ and suppose it comprises
nodes $v_{k+1},v_k, v_{k-1}, \ldots, v_1$
 from bottom to top ($v_1$ is
the root of the join tree and 
$v_{k+1}$ is the join node that is the parent of the final node in the stolen subtask).

Let $P$ be the processor executing $\tau$ initially
and let $P_1$ be the processor executing 
stolen task
$\tau_1$.
Suppose the timing is such that $P_1$ executes
all the nodes on $\overline{\cal P}$.
Then, $P$ executes 
$k= \Theta (\log n)$
 non-contiguous fragments of sequential computation, one for each $\mu_i$.
Likewise $P_1$ executes each of the $v_i$ in turn, and these are also non-contiguous.
Since $\Theta(\log n)$ fragments are created by one steal,
 the simple argument providing the $O(M/B)$ cache miss overhead
per steal will not apply to general schedulers.
}
\end{example}

\def\mo{\mu_1}
\def\mw{\mu_2}
\def\mr{\mu_3}
\def\to{\tau_1}
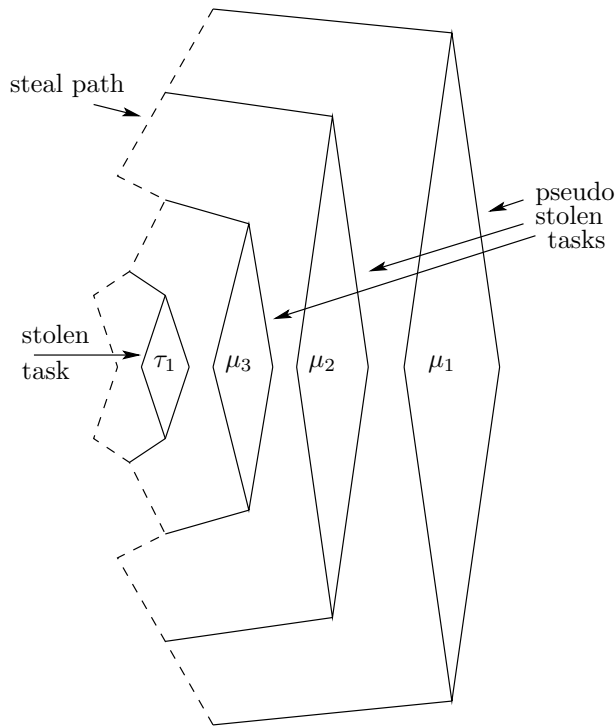
\begin{figure}[htbp]
\begin{center}

\begin{picture}(0,0)%
\includegraphics{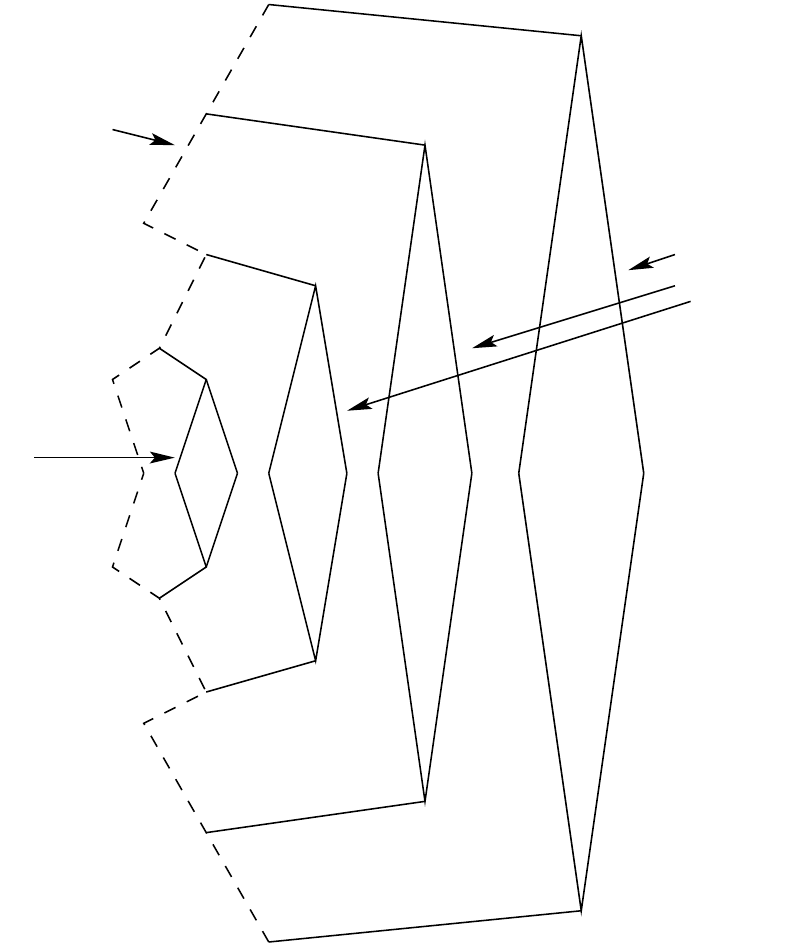}%
\end{picture}%
\setlength{\unitlength}{3947sp}%
\begingroup\makeatletter\ifx\SetFigFont\undefined%
\gdef\SetFigFont#1#2#3#4#5{%
  \reset@font\fontsize{#1}{#2pt}%
  \fontfamily{#3}\fontseries{#4}\fontshape{#5}%
  \selectfont}%
\fi\endgroup%
\begin{picture}(3760,4524)(1111,-3673)
\put(1201,-1486){\makebox(0,0)[lb]{\smash{{\SetFigFont{10}{12.0}{\rmdefault}{\mddefault}{\updefault}{\color[rgb]{0,0,0}task}%
}}}}
\put(1201,-1261){\makebox(0,0)[lb]{\smash{{\SetFigFont{10}{12.0}{\rmdefault}{\mddefault}{\updefault}{\color[rgb]{0,0,0}stolen}%
}}}}
\put(1126,314){\makebox(0,0)[lb]{\smash{{\SetFigFont{10}{12.0}{\rmdefault}{\mddefault}{\updefault}{\color[rgb]{0,0,0}steal path}%
}}}}
\put(3751,-1411){\makebox(0,0)[lb]{\smash{{\SetFigFont{10}{12.0}{\rmdefault}{\mddefault}{\updefault}{\color[rgb]{0,0,0}$\mo$}%
}}}}
\put(3001,-1411){\makebox(0,0)[lb]{\smash{{\SetFigFont{10}{12.0}{\rmdefault}{\mddefault}{\updefault}{\color[rgb]{0,0,0}$\mw$}%
}}}}
\put(2476,-1411){\makebox(0,0)[lb]{\smash{{\SetFigFont{10}{12.0}{\rmdefault}{\mddefault}{\updefault}{\color[rgb]{0,0,0}$\mr$}%
}}}}
\put(2026,-1411){\makebox(0,0)[lb]{\smash{{\SetFigFont{10}{12.0}{\rmdefault}{\mddefault}{\updefault}{\color[rgb]{0,0,0}$\to$}%
}}}}
\put(4501,-661){\makebox(0,0)[lb]{\smash{{\SetFigFont{10}{12.0}{\rmdefault}{\mddefault}{\updefault}{\color[rgb]{0,0,0}tasks}%
}}}}
\put(4426,-511){\makebox(0,0)[lb]{\smash{{\SetFigFont{10}{12.0}{\rmdefault}{\mddefault}{\updefault}{\color[rgb]{0,0,0}stolen}%
}}}}
\put(4426,-361){\makebox(0,0)[lb]{\smash{{\SetFigFont{10}{12.0}{\rmdefault}{\mddefault}{\updefault}{\color[rgb]{0,0,0}pseudo}%
}}}}
\end{picture}%

\caption{\label{fig:pseudo-stolen}Illustrating Example~\ref{ex:gen-sched-interleave}.
Here there are four stealable tasks to the right of the steal path, $\mu_1, \mu_2, \mu_3, \tau_1$.
With a general scheduler, if the lowest such
subtask, $\tau_1$, were the only one that was stolen, 
this would be a deep steal inducing a pseudo task kernel
consisting of the three other stealable
tasks but not the portions of the steal path
connecting them.}

\end{center}
\end{figure}

In the next section we recover the $R(S)=O(S \cdot M/B)$ bound for a general scheduler
 (for all series parallel computation dags) using a different
 analysis, and in Section~\ref{sec:general} we present a further refined analysis for HBP algorithms.

%% file: refined-summary.tex
\section{The New Cache Miss Analysis}
\label{sec:cache-summary}

We begin by specifying the partitioning of the computation into 
{\it task kernels}
in Section~\ref{sec:task-kernels}.
We follow this by demonstrating 
in Section~\ref{sec:thm1}
the simple $O(Q+ S\cdot M/B)$ bound on the cache miss cost.
We then outline our more sophisticated bound, analyzing in turn 
BP computations in Section~\ref{sec:bp} and HBP computations in Section~\ref{sec:hbp-outline}.

\subsection{Tasks and Task Kernels}\label{sec:task-kernels}

Let $\tau$ be a task that incurs steals.
Informally, a task kernel of $\tau$ is a maximal contiguous fragment of the computation that
lies entirely within the unstolen portion of $\tau$ or 
entirely
within a stolen task,
and represents a connected sub-dag when
the computation dag is partitioned at each join node on the steal path
immediately following the end of a stolen subtask.

We now define the task kernels induced by the steals as follows.
This definition applies to any computation dag in which any given pair of
forks and joins is either
nested or disjoint, as for example in series-parallel dags,
and this includes all
HBP computations. 
Figure~\ref{fig:task-kernel-bp} gives examples of task kernels in a BP computation.

\begin{definition}
\label{def:task-kernel-general}
{\bf  (Task kernels.)}
Consider a parallel execution of a computation $C$ under a general scheduler, 
and suppose it incurs $S$ steals,
 $\sigma_1, \sigma_2 , \cdots , \sigma_S$,
numbered 
in the order
in which the stolen tasks are generated (i.e., 
the order in which the parent fork nodes are executed)
in a sequential execution.
In turn, we partition the computation dag
into task kernels with respect to the sequence 
$\Sigma_i  = \langle \sigma_1, \sigma_2, \ldots, \sigma_i \rangle$
to create the collection $C_i$.
We let $\Sigma_0$ be the empty sequence and its associated partition $C_0$ be the
single task kernel containing the entire computation dag.
For each $i \geq 1$ the partition $C_{i+1}$ is obtained from $C_i$
as follows. 
Let $\tau$ be the task kernel in $C_i$
that contains the fork node $v_f$ at which steal $\sigma_{i+1}$ is performed, 
and let $v_j$ be the corresponding join node. 
Then, $\tau$ is partitioned into the following  task kernels, each 
categorized as being of type \emph{starting, finishing,} or \emph{pseudo}.
The initial task kernel 
in $C_0$
is given type starting.

1. $\tau_1$, the stolen subtask created by $\sigma_{i+1}$.
It is called
a \emph{starting task kernel.}

2. $\tau_2$, the  portion of $\tau$ preceding the stolen subtask in the sequential execution
(this includes the portion of the computation descending from the left child of $v_f$
that precedes $v_j$.)
It is given
 the same type as $\tau$.

3. If $\sigma_{i+1}$ is a deep steal, let $\nu =\langle \mu_1, \mu_2, \ldots, \mu_k\rangle$ be the sequence of
pseudo-stolen tasks forked from $\tau$  that are on the task queue for the processor
executing $\tau$ at the time of this steal, with $\mu_k$ immediately
preceding the subtask stolen by $\sigma_{i+1}$.
Suppose that $\mu_{i_1}, \mu_{i_2}, \ldots, \mu_{i_j}$ 
(for $i_1 < i_2 < \cdots$)
incur steals
(note that any such steal would occur after $\sigma_{i+1}$ in our ordering of steals).
Then each of the collections 
$(\mu_1, \mu_2, \ldots, \mu_{i_1 -1})$, $(\mu_{i_1}, \mu_{i_1+1},\ldots, \mu_{i_2-1})$,
$\ldots$, $(\mu_{i_j}, \mu_{i_j+1}, \ldots, \mu_k)$ forms a \emph{pseudo task kernel}.

4. $\tau_3$, the portion of $\tau$ 
starting at $v_j$ in the sequential execution but excluding 
the collection $\nu$ in part 3 above.
This includes the join nodes following the pseudo-stolen tasks in 
$\nu$.
Then, $C_{i+1} = (C_i - \{\tau\}) \cup \{ \tau_1, \tau_2, \tau_3\}
\cup \{\mbox{pseudo task kernels}$
$\mbox{~formed in part 3, if any}\}$.

The final collection $C_S$ is the collection of task kernels for this parallel execution of $C$.
\end{definition}

\def\vf{v_f}
\def\vj{v_j}
\def\tt{\tau_2}
\def\tl{\tau_l}
\begin{figure}[htbp]
\begin{center}

\begin{picture}(0,0)%
\includegraphics{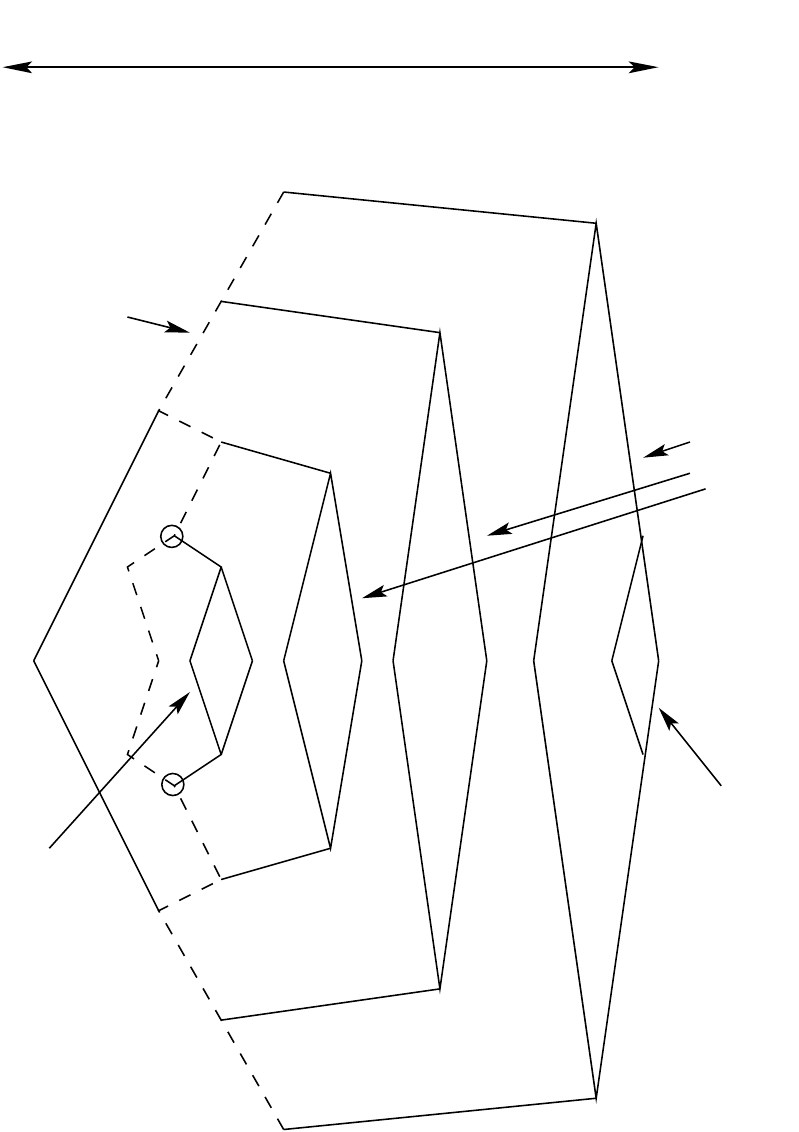}%
\end{picture}%
\setlength{\unitlength}{3947sp}%
\begingroup\makeatletter\ifx\SetFigFont\undefined%
\gdef\SetFigFont#1#2#3#4#5{%
  \reset@font\fontsize{#1}{#2pt}%
  \fontfamily{#3}\fontseries{#4}\fontshape{#5}%
  \selectfont}%
\fi\endgroup%
\begin{picture}(3832,5421)(1039,-3673)
\put(1126,-2611){\makebox(0,0)[lb]{\smash{{\SetFigFont{10}{12.0}{\rmdefault}{\mddefault}{\updefault}{\color[rgb]{0,0,0}task}%
}}}}
\put(4351,-2161){\makebox(0,0)[lb]{\smash{{\SetFigFont{10}{12.0}{\rmdefault}{\mddefault}{\updefault}{\color[rgb]{0,0,0}stolen}%
}}}}
\put(1126,-2461){\makebox(0,0)[lb]{\smash{{\SetFigFont{10}{12.0}{\rmdefault}{\mddefault}{\updefault}{\color[rgb]{0,0,0}stolen}%
}}}}
\put(1126,314){\makebox(0,0)[lb]{\smash{{\SetFigFont{10}{12.0}{\rmdefault}{\mddefault}{\updefault}{\color[rgb]{0,0,0}steal path}%
}}}}
\put(3001,-1411){\makebox(0,0)[lb]{\smash{{\SetFigFont{10}{12.0}{\rmdefault}{\mddefault}{\updefault}{\color[rgb]{0,0,0}$\mw$}%
}}}}
\put(2476,-1411){\makebox(0,0)[lb]{\smash{{\SetFigFont{10}{12.0}{\rmdefault}{\mddefault}{\updefault}{\color[rgb]{0,0,0}$\mr$}%
}}}}
\put(2026,-1411){\makebox(0,0)[lb]{\smash{{\SetFigFont{10}{12.0}{\rmdefault}{\mddefault}{\updefault}{\color[rgb]{0,0,0}$\to$}%
}}}}
\put(4501,-661){\makebox(0,0)[lb]{\smash{{\SetFigFont{10}{12.0}{\rmdefault}{\mddefault}{\updefault}{\color[rgb]{0,0,0}tasks}%
}}}}
\put(4426,-511){\makebox(0,0)[lb]{\smash{{\SetFigFont{10}{12.0}{\rmdefault}{\mddefault}{\updefault}{\color[rgb]{0,0,0}stolen}%
}}}}
\put(4426,-361){\makebox(0,0)[lb]{\smash{{\SetFigFont{10}{12.0}{\rmdefault}{\mddefault}{\updefault}{\color[rgb]{0,0,0}pseudo}%
}}}}
\put(3676,-1636){\makebox(0,0)[lb]{\smash{{\SetFigFont{10}{12.0}{\rmdefault}{\mddefault}{\updefault}{\color[rgb]{0,0,0}$\mo$}%
}}}}
\put(4051,-1411){\makebox(0,0)[lb]{\smash{{\SetFigFont{10}{12.0}{\rmdefault}{\mddefault}{\updefault}{\color[rgb]{0,0,0}$\tt$}%
}}}}
\put(1801,1589){\makebox(0,0)[lb]{\smash{{\SetFigFont{12}{14.4}{\rmdefault}{\mddefault}{\updefault}{\color[rgb]{0,0,0}Stolen task $\tau$}%
}}}}
\put(1426,-1561){\makebox(0,0)[lb]{\smash{{\SetFigFont{12}{14.4}{\rmdefault}{\mddefault}{\updefault}{\color[rgb]{0,0,0}$\tl$}%
}}}}
\put(1801,-511){\makebox(0,0)[lb]{\smash{{\SetFigFont{12}{14.4}{\rmdefault}{\mddefault}{\updefault}{\color[rgb]{0,0,0}$\tl$}%
}}}}
\put(1651,-2086){\makebox(0,0)[lb]{\smash{{\SetFigFont{12}{14.4}{\rmdefault}{\mddefault}{\updefault}{\color[rgb]{0,0,0}$\vj$}%
}}}}
\put(1726,-2386){\makebox(0,0)[lb]{\smash{{\SetFigFont{12}{14.4}{\rmdefault}{\mddefault}{\updefault}{\color[rgb]{0,0,0}$\tl$}%
}}}}
\put(1651,-736){\makebox(0,0)[lb]{\smash{{\SetFigFont{12}{14.4}{\rmdefault}{\mddefault}{\updefault}{\color[rgb]{0,0,0}$\vf$}%
}}}}
\put(4351,-2311){\makebox(0,0)[lb]{\smash{{\SetFigFont{10}{12.0}{\rmdefault}{\mddefault}{\updefault}{\color[rgb]{0,0,0}task}%
}}}}
\end{picture}%

\caption{\label{fig:task-kernel-bp} Task kernel types in a BP computation.
Suppose a task 
$\tau$ undergoes a deep steal of subtask $\tau_1$ at fork node $v_f$.
Then $\tau_1$ forms a starting task kernel.
Also, $\tau_l$,
the portion of $\tau$ to the left of the steal path, plus the steal path up to but
not including $v_j$, forms another starting  task kernel.
$\mu_1, \mu_2, \mu_3$ are all pseudo stolen tasks;
if $\mu_1$ undergoes a steal but $\mu_2$ and $\mu_3$ do not, 
then $\mu_2 \cup \mu_3$ form a pseudo task kernel.
Finally, the portion of the path of join nodes
descending from $v_j$, and including $v_j$ forms a finishing task kernel.
(A finishing task kernel can incur a steal only in 
a Type $k$ HBP, for $k>1$.)
}
\end{center}
\end{figure}
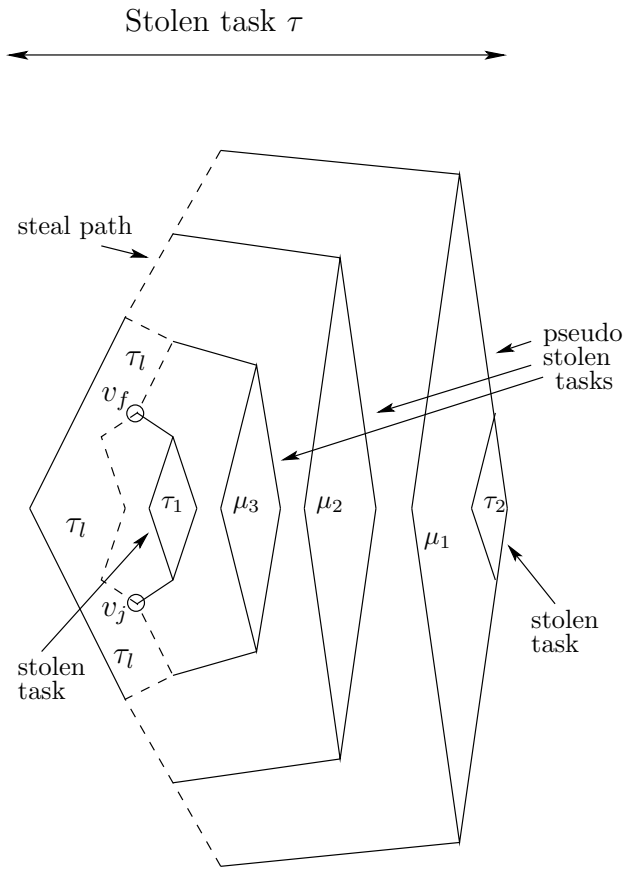

In part 3 above, each  pseudo task kernel comprises a maximal sequence of pseudo-stolen task kernels, which in execution
order ends
 with a pseudo-stolen kernel that incurs a steal, with all the other pseudo-stolen task kernels in the
sequence being steal-free. In part 4, the finishing task kernel $\mu_3$ comprises the nodes 
descendant from $v_j$ in the computation dag, including $v_j$ itself.
Note that in the sequential execution,
both the 
finishing
task kernel $\mu_3$ and the 
pseudo task kernels in $\nu$ are
executed after the stolen subtask,
and they interleave in their execution.
Some implications of this interleaving were explored in Example~\ref{ex:gen-sched-interleave}.

\begin{lemma}\label{lem:task-kernel-number}
A series parallel computation with $S$ steals has at most $4S+1$ 
task kernels,
of which at most $S+1$ are starting kernels, $S$ are finishing, and $2S$ are pseudo.
\end{lemma}
\begin{proof} 
In the absence of a deep steal, the number of task kernels is 
exactly $2S+1$, since there is initially one task kernel, and each successive steal replaces
a current task kernel with three new ones, according to parts 1, 2 and 4 in 
Definition~\ref{def:task-kernel-general}, 
one finishing, one starting, and one of the previous type.
This yields at most $S+1$ starting and $S$ finishing task kernels.

Now, let us consider the effect of part 3 in Definition~\ref{def:task-kernel-general}, which
creates $i_{j}+1$ pseudo task kernels when $i_j$ of the pseudo stolen tasks in $\nu$ incur
steals. We claim that we can bound the number of pseudo task kernels 
by
 $2S$ 
by charging at most two of them to each steal as follows. To each deep steal
 $\sigma$ we assign the
last pseudo task kernel $(\mu_{i_j+1}, \mu_{i_j+2}, \ldots, \mu_k)$ in its collection $\nu$ 
(as defined in part 3 of Definition~\ref{def:task-kernel-general}). We assign each of the remaining
pseudo task kernels for $\sigma$ to the earliest steal  $\sigma'$ (in our ordering) in the steal-incurring pseudo-stolen task kernel. Now consider $\sigma'$. It may be assigned 
 another
pseudo task kernel if it is itself a deep steal (in a different state of the execution stack). So 
$\sigma'$ could be assigned
 two different pseudo task kernels. But it cannot be assigned 
a third one, since for any pseudo stolen task that contains $\nu$, the earliest steal in it is
either $\sigma$ or a steal earlier than $\sigma$.
 Thus, there are at most $2S$ additional
task kernels created due to the pseudo task kernels, and this adds up to a total of at most
$4S +1$ task kernels.
\end{proof}

\begin{ignore}
We observe that with a general scheduler,
a computation incurring $S$ steals can have up to $4S+1$ task kernels
since $C_{i+1}$ is obtained from $C_i$ by partitioning a task kernel in $C_i$ into up to four new task kernels
plus $j_{i+1}$ additional pseudo task kernels in the event that $j_{i+1}$ of the pseudo-stolen task kernels induced
by steal $\sigma_{i+1}$ are steal incurring; the key point to note is that each steal can induce
only one additional pseudo task kernel.
Also, as noted in Section~\ref{execstack-general}, additional execution stacks may be needed 
to execute a pseudo-stolen task kernel when using a general scheduler.
\end{ignore}

\subsection{Proof of Theorem 1}\label{sec:thm1}

Recall Example~\ref{ex:gen-sched-interleave}.
Observe that the
sequence
$\langle \mu_k, v_k, \mu_{k-1}$, $v_{k-1},$
$\ldots, \mu_1, v_1\rangle$
  is contiguous.
Thus each of $P_1$ and $P$ executes a portion of the same contiguous
sequence, and between them they execute all of it.
Therefore their combined 
caching overhead is at most twice the sequential cost
plus an additional $M/B$ term for each of them.
The proof of Theorem~\ref{thm:cache-miss-simple} will 
build on this insight to
 establish that in fact 
$C(S) = 2Q +O(S\cdot M/B)$
under a general scheduler for the entire class
 of series-parallel dags.

\begin{proof}[Proof of Theorem~\ref{thm:cache-miss-simple}]
For the purposes of this proof, we further refine the partitioning into task kernels as follows.
Let $\mu$ be a pseudo task kernel that ends in a steal-incurring pseudo-stolen task.
Let $v_t$ be the terminal node in $\mu$, i.e., the final node in $\mu$ in a sequential execution.
Let $v_j$ be
 the node following $v_t$ in a sequential execution;
then $v_j$ is a join node and it
 lies in a finishing
task kernel, which we will denote by $\nu^{\mu}$. 
We split $\nu^{\mu}$ in two, where the
initial portion $\nu^{\mu}_1$ contains the portion of $\nu^{\mu}$ up to, but not including
$v_j$, and the latter portion $\nu^{\mu}_2$ contains the portion of $\nu^{\mu}$ starting at
$v_j$. We then merge $\mu$ with $\nu^{\mu}_1$ to form a
{\it super-finishing task kernel} $\mu$-$\nu$ (and we discard  $\mu$ and $\nu^{\mu}_1$). 
We repeatedly perform this split
and merge into super-finishing task kernels
at each steal-incurring pseudo task kernel. 

The above process partitions the computation
into at most $4S+1$
task kernels, 
some of which may be super-finishing task kernels.
 Each of these task kernels
has the 
useful
property that it
is executed contiguously in a sequential execution.
Furthermore, at most two processors are used to execute each super-finishing task kernel, namely the processor
starting the corresponding finishing kernel and the processor completing the corresponding pseudo task kernel.
The other task kernels are all executed by a single processor.

Thus, as in the work-stealing case, 
for each of these $4S+1$
kernels there is a cost of $O(M/B)$ cache misses to restore the state that existed in the sequential execution, 
and as the execution of 
each of the at most $S$
 super-finishing 
kernels
is shared among two processors, 
this at most doubles the cache miss cost of these portions of the computation,
leading to a bound of at most  $2Q+(5S +1) \cdot M/B$
 additional cache misses due to the
steals. 
This establishes the desired bound.
\end{proof}

%% file: general-overview.tex
\section{The HBP Analysis}\label{sec:general}

In this section we present an improved bound on
 the caching overhead of HBP algorithms under a general
scheduler. Our approach to improving the bound in Theorem~\ref{thm:cache-miss-simple} is to carefully
examine the features of HBP algorithms and tailor our analysis to algorithms in this class.

\subsection{Reload Cost}\label{sec:reload}

We now define the notion of the reload cost of a sequence of steps executed within a sequential execution of a task $\tau$.

\begin{definition}\label{def:reload}
{\bf (Reload Cost)}
Let $\mu$ be a sequence of step within a task $\tau$ that are executed contiguously in a sequential execution of $\tau$. 
The \emph{reload cost of $\mu$} is the number of distinct blocks accessed by $\mu$ during its execution,
excluding blocks that contain variables declared during $\mu$'s computation.
\end{definition}

In our analysis we will use the reload cost in place of the simple upper bound of $M/B$ for the additional cache miss cost in
executing a stolen task, or any task kernel that consists of the steps executed contiguously in a sequential execution (typically
starting task kernels). We
will use the following lemma in our analysis.

\begin{lemma}\label{lem:reload-cost}
Let $\mu$ be a sequence of step within a task $\tau$ that are executed contiguously in a sequential execution of $\tau$. 
Let $Q$ be an upper bound on the number of cache misses incurred by
$\tau$ during its execution of $\mu$ in a sequential execution. If $\mu$ is executed as a separate computation, then its cache miss cost is
$O(Q+R)$, where $R$ is its reload cost.
\end{lemma}

\begin{proof}
Let us consider the additional cache miss cost in a separate execution of $\mu$ for
reading in data that may have already resided in cache in an execution of $\mu$ within an execution of $\tau$.
Let us refer to the variables accessed by the execution of $\mu$ excluding variables declared during $\mu$'s
computation as
{\it new variables}, and the $R$ blocks in memory in which they reside as {\it new blocks}.
The only difference between a separate execution of $\mu$ and the execution of $\mu$ within a sequential execution of $\tau$ is that some
of the $R$ new blocks may already be in cache at the start of the latter execution, and hence the cost of reading these blocks is not included in $Q$.
Now consider a separate execution of $\mu$.
 If $\mu$ does not evict any of the $R$
new blocks during its separate execution, then its cache miss cost is bounded by $O(Q + R)$ since the two executions only differ in the initial presence of these $R$ blocks. On the other hand, any new block evicted by $\mu$ in its separate execution must also be evicted by $\tau$ in its execution of $\mu$ since both perform the same computation and both are assumed to use a given optimal cache replacement policy.
Hence the number of cache
misses
in a separate execution of $\mu$ is bounded by $O(Q+R)$.
\end{proof}

The above proof does not address the `block misalignment' cost~\cite{CR10} that arises from the fact that the block boundaries for data  on an execution stack may be different in a parallel execution of a task from
what they
would be in a sequential execution. However, it is shown in~\cite{CR10} that its effect
 is
bounded by a constant factor for HBP computations, 
if the cache miss bound is polynomial in $M$ and $B$\footnote{The polynomial dependence on $M$ and $B$
is implicit in the earlier work.},
so the bound in the above lemma holds even when accounting for
block misalignment costs.
In fact, this observation extends to the full analysis of the HPB algorithms in this paper.

\subsection{Data Layout and Caching Costs}\label{sec:data}

The caching cost of a computation, even in the sequential context, is highly dependent on the
 data layout and the pattern of accesses to the data during the computation. Since we are
 bounding the caching overhead for a class of  algorithms, rather than for a specific algorithm, we
 now specify the type of data layouts that we allow in the algorithms we analyze, and the data access
 patterns. We focus mainly on BP computations, since that is where the most of variation in data
 layout occurs. An HBP computation may declare shared arrays which are accessed by  BP computations
  within its recursive computations.
 
A BP computation will access access variables placed on the execution stack by its fork 
and join
nodes as well as the shared data structures declared at the start of the computation. Recall that
each BP node performs $O(1)$ computation. Here are the types of accesses we allow in the
algorithms for which we bound
the cache miss overhead.

\vhalf
\noindent
{\it Accesses to the Execution Stack.}  A BP or HBP node can access its $O(1)$ data, and it can also access
data declared by its parent node.

\vhalf
\noindent
{\it Accesses to Shared Data.}
Consider a shared array where each data item is associated with a single node in the BP tree.
Depending on the algorithm, a node may access just its associated data, or its data plus the
data for some or all of its neighbors (children and parent). We will assume 
 that the data in this array 
 is laid out contiguously according to an {\it inorder} traversal of the BP fork tree. Our results will go through
 if preordering or postordering is used instead of the inorder traversal we assume. We choose to use
 inorder traversal because it aids in obtaining our false sharing results (false sharing is mentioned in the introduction, but is not included here).
 
 Recall that we refer to the computation dag as a task, and any parallel task spawned by a fork node is
 also a task. We now define the notion of an extended task. 
Here we use the convention that a fork tree includes the non-fork leaf nodes that lie between it 
 and the complementary join tree.
 
 \begin{definition}\label{def:inorder}
{\bf (Extended task)}
 An \emph{extended task} in a BP computation
is any sequence of $r>1$ consecutive nodes in the inorder traversal of its fork tree
together with the complementary join nodes.
\end{definition}

Any task is clearly  also an extended task.
From the definition of a starting task kernel, we
 can see that it is basically an extended task, except that some
 of the nodes on its steal path may not lie
 within this extended task kernel. Our cache miss analysis will separately analyze the costs due to the
 portion of the starting task kernel that forms an extended task, and the portion outside of this 
 extended task.

We now define the {\it data dispersal} function 
$f(r)$
(previously called the cache friendliness function in \cite{CR12b}), 
 which 
parameterizes the cost of  accesses to the shared data structures.

\begin{definition}\label{def:fr}
\label{def:f-r}
{\bf (Data dispersal function $f(r)$ for BP computations)}
A collection of $r$ words is $f(r)$-dispersed
if it is contained in $(r/B) + f(r)$ blocks.
An extended
 task is $f(r)$-dispersed
if the data
its nodes access when executed is contained in $(r/B) + f(r)$ blocks.
A BP
computation is $f(r)$-dispersed
 if every extended
 task in it is $f(r)$-dispersed.
\end{definition}

Notions similar to our use of $f(r)$ have been used in 
sequential 
caching analyses, though 
our set-up is more general,
and this generality is needed  to obtain bounds for  the general
class of  BP and HBP computations, as opposed to analyzing a single algorithm.

Examples of data dispersal functions for algorithms for scans, prefix sums and matrix transpose are
given
 in Appendix~\ref{sec:f-r-examples}.
The algorithms 
in Table~\ref{table1}
(except for two procedures in SPMS sorting)
are all 
$O(\sqrt r)$-dispersed, and 
the ones for scans and prefix sums can be made $O(1)$-dispersed
(for which we give an improved bound in Section~\ref{sec:bp}).
As a result they satisfy our assumption in Section~\ref{sec:results} that a task that 
accesses $r$ words
will access $O((r/B) + \sqrt r)$ blocks. We present this more general analysis here 
using $f(r)$
since it allows one 
to fully analyze the SPMS algorithm and other algorithms with complex data access patterns.

%% file: bp-general.tex
\subsection{BP and Type 1 HBP Computations}\label{sec:bp}

Consider a BP computation $\tau$. We 
begin with a high-level description of the structure
of task kernels in a BP computation, which are also illustrated in Figure~\ref{fig:task-kernel-bp}.

\vhalf
\noindent
{\bf Starting task kernel.}
We first observe that in a BP computation, a
starting task kernel will consist of a zig-zag path in the fork tree
(the steal path)
 with subtasks comprising its off-path
left subtrees, together with the complementary subtrees in the join tree, but not the complementary zig-zag path
in the join tree, 
for it forms a finishing task kernel. Each left-going segment in the 
fork tree
zig-zag path contains the parents of stolen
or pseudo-stolen tasks, and the left subtrees in each right-going segment are part of the starting task kernel, 
as implicitly specified
in Definition~\ref{def:task-kernel-general}.

\vhalf
\noindent
{\bf Pseudo task kernel.}
A pseudo task kernel 
 comprises a sequence of one of more pseudo-stolen tasks,
where each pseudo-stolen task is itself a 
(smaller)
BP computation which returns to a parent node on the join path of the
 task from which it was ``stolen.''
 This join node is not part of the pseudo-stolen task or the pseudo task kernel.
The topmost
pseudo-stolen task in the pseudo task kernel
 may have incurred steals and as a result
may have the same form as for a starting
task kernel. The remaining pseudo-stolen tasks, if any,
 are steal-free.

\vhalf
\noindent
{\bf Finishing task kernel.}
In a BP computation,
a finishing task kernel is 
simply
a path in the join tree that
ends at a parent of a returning stolen task, or at the root.

\vone
\noindent
{\bf Accessing the Shared Data Structures.} 
We start by
 analyzing the cost of accesses to the shared data structures, and we
 first analyze this
 cost for finishing kernels.
As noted above, in a BP computation, each finishing task kernel 
is simply a path in the join tree, and these paths
are disjoint. As the computation at a node may also access the data for its parent,
the accesses by 
a finishing task kernel may overlap 
those by other task kernels, but only at their
 end nodes.

We will bound the overall cache miss cost for all finishing task kernels, by separately analyzing
 the cost of accesses 
to nodes in 
the topmost $\log |\tau| - \log B$ levels in the join tree, and 
then
to the nodes in 
the bottom $\log B$ levels. 
There are only  $|\tau|/B$ nodes in total in the 
topmost $\log |\tau| - \log B$ levels, and each causes
$O(1)$ cache misses and so collectively they
have a cost of no more than $O(|\tau|/B)$ cache misses;
this cost could be smaller if the $O(S)$ finishing
task kernel join paths traverse fewer than $\Theta(|\tau|/B)$ nodes in 
this portion of the join tree.
There are $O(\log B)$ accesses by each
finishing task kernel to nodes in the bottom $\log B$ levels
of the join tree;
we simply charge 
these $O(\log B)$ accesses to each of these task kernels, 
which adds up to $O(S \cdot \log B)$ cache misses across all finishing task kernels. Hence
the total cost of cache misses for finishing task kernels is $O(|\tau|/B + S \cdot \log B)$.

In the case that $f(r) = O(1)$ we can improve the $O(S\cdot \log B)$ term to $O(S)$.
The reason is that the nodes in the subtree of height $\log B$ comprise
$O(B)$ contiguous nodes in the inorder traversal 
and hence the data these $B$ nodes access is stored in $O(1 +f(1)) = O(1)$ blocks.
Clearly this bound also applies to the subset of $\log B$ nodes on the path in question.
Consequently the
execution of all these nodes, for each path, will incur $O(1)$ cache misses,
or $O(S)$ cache misses when summed over all the finishing task kernels.
This yields 
a total cost of
$O(|\tau|/B + S)$ cache misses.

\vhalf
For a starting task kernel, 
the nodes on the zig-zag path in the fork tree
which have off-path left children, plus their off-path subtrees,
together with the complementary subtrees, 
are contiguous in inorder, and hence form an extended task.
However, the accesses by 
the nodes on the remainder of the fork tree
zig-zag path,
namely the nodes with off-path right subtrees,
will be non-contiguous.
To bound this cost, we observe that
 the cost of the shared data structure accesses by these nodes 
is no larger than the bound for the finishing task kernels,
as the complement of each fork tree zig-zag path is 
the union of one or more finishing task kernels.
Hence we add in the 
charged cost to the complementary finishing task kernels.
Clearly, each finishing task kernel is charged at most once.

\vhalf
In a pseudo task kernel $\mu$, the data accesses 
are to the data for the nodes
in the kernel plus the data for the parents of pseudo-stolen tasks forming  $\mu$,
and this is a contiguous collection of nodes in the inorder traversal, 
aside any additional discontinuities caused by steals from $\mu$, if any; 
any steals from the pseudo task kernel $\mu$ cause the same sort of discontinuities
as the steals from a starting kernel, as discussed in the previous paragraph,
and are bounded 
by a similar charging scheme.

\vhalf
Aside the accesses to the data for the portions of the fork tree
zig-zag path nodes specified two paragraphs above, 
we see that every starting and pseudo task kernel
is accessing a disjoint contiguous interval of nodes
in inorder, hence by Definitions~\ref{def:reload} and \ref{def:f-r} and Lemma~\ref{lem:reload-cost},
the cache miss cost for these is bounded by
$O(\sum_{i=1}^k r_i/B + f(r_i))$, where 
 $r_i$ is the number of nodes 
in the fork tree  that are also
in the $i$-th starting or pseudo kernel,
$\sum_{i=1}^k r_i= n$,
and there are $k= O(S)$ of these kernels in total.
This sum totals $O(n/B + \sum_{i=1}^k  f(r_i))$.

We now give tight upper bounds on this term for $f(r) = O(1)$ and $f(r) = O(\sqrt r)$.
Clearly, when $f(r) = O(1)$ this term is just $O(n/B + S)$. For
$f(r) =O( \sqrt r)$, 
the sum $\sum_{i=1}^k f(r_i)$ is maximized when the $r_i$ are all equal, and contributes the term
$O(S\sqrt{n/S}) = O(\sqrt{Sn})$.
When $S \le n/B^2$ this is $O(n/B)$ and when $S > n/B^2$, this is $O(S\cdot B)$.
Thus it is always bounded by $O(n/B + S \cdot B)$.

 \vone
 \noindent
 {\bf Accessing the Execution Stacks.} 
 The
overhead at execution stacks occurs when a stolen 
or pseudo-stolen
task
 accesses the execution stack for its parent task when it returns from its computation, and
 when the subsequent finishing task kernel 
(in the case of a stolen task)
 has to reload the segments on the execution stack that
it needs to access. 
This entails $O(1)$ accesses by each stolen 
task,
 $O(1)$ accesses by each pseudo-stolen task
to the data segment for a distinct node in a finishing task 
kernel since
each pseudo-stolen task is accessing its
parent in the join tree,
 and $O(1)$ accesses by each node in each finishing task
kernel, since each finishing task kernel accesses $O(1)$ data on the execution stack at each node
 on the join path that comprises the finishing task kernel.
Since the segments for the nodes in each finishing task are consecutive on its execution stack,
the cost for accessing a path of $l$ nodes is $O(\ceil{l/B})$ cache misses.
Furthermore, each 
pseudo task kernel
 will be accessing a portion of a zig-zag path
in the join tree, and it will be the only 
pseudo task kernel to access
each of these nodes, aside from one finishing task kernel.
Thus the cost of the accesses
to the parent nodes in the join tree by each 
pseudo task kernel
is no more that the cost for the corresponding finishing task kernel.

The total length of the paths for the finishing tasks
is $O(|\tau|)$. Thus the cost of the accesses to the segments on the execution stack is
bounded by $O(|\tau|/B + S)$;
The $+S$ term is due to the rounding up of the term for each of the $O(S)$ paths,
one path for each of the $O(S)$ finishing task kernels.
Finally, the accesses by the stolen tasks add
 another $O(S)$ accesses to the total, 
which therefore sums to  $O(|\tau|/B + S)$ cache misses.
This bound for accesses to the execution stacks, together with the earlier bound for accesses by the task kernels, leads
to the following lemma.

\begin{lemma}\label{lem:bp}
Consider a BP computation of size $n$.
When scheduled with a general scheduler, it will incur $O((n/B) + S\cdot B)$ cache misses
if $f(r) = O(\sqrt r)$
and $O(n/B + S)$ cache misses if $f(r) = O(1)$, where $S$ is the number of steals.
In general, the number of cache misses will be bounded by 
$O(n/B + \max \sum_{i=1}^{3S+1} f(r_i))$ where $\sum_{i=1}^{3S+1} r_i = n$.
\end{lemma}
\begin{proof}
The expression for the general bound arises because 
by Lemma~\ref{lem:task-kernel-number},
there are at most 
$3S+1$ starting and pseudo kernels, each of which contributes 
to at most one of the $f(r_i)$ terms.
\end{proof}
This establishes part $(i)$ of Theorem~\ref{thm:cache-miss-overhead-gen}.

Note that it is only the accesses to the shared data structure that depend on the function $f(r)$.
The bound for the accesses to the execution stacks is always $O(|\tau|/B + S)$.
Other more irregular patterns of access 
to the shared data structure
can arise and these would need to be analyzed separately.

\vhalf
\noindent
{\bf Type 1 HBP Computations}
The one extra feature in a Type 1
HBP computation is that if one constituent $\mu$ of the computation incurs
a steal then the finishing task kernel that emerges from $\mu$ may need to access variables previously accessed
in the computation of $\mu$ (or earlier constituents, if any). But this entails $O(|\tau|/B)$
cache misses,
and occurs at most once for each constituent (except the first), which is a total additional cost of
$O(|\tau|/B)$ cache misses, as each Type 1
algorithm has $O(1)$ constituents
by Definition~\ref{def:hbp}
(and this is the reason for this restriction).
This leads to the bound in Table 1 for Prefix Sums. 
Also, it turns out that a more careful analysis shows that the second prefix sums algorithm
described earlier, which had $f(r) = O(\log (r/B))$,
achieves the same bound as the algorithm
with $f(r) = O(1)$. We omit the details here.

%% file: hbp-general.tex
\subsection{Outline of the HBP Analysis}
\label{sec:hbp-outline}

We give here an overview of the HBP analysis. 
The full details, and several proofs, are deferred to the Appendix.

We start by defining the notions of fork tree ownership
and local and remote steals.
 
\begin{definition}
\label{def:fork-tree-own}
An HBP task $\tau$ owns the fork trees for its constituent tasks.
A steal in $\tau$ that occurs in a fork tree it owns is a \emph{local steal} and is owned by $\tau$.
A steal in $\tau$ that occurs in a recursive task and not in a fork tree it owns is called
a \emph{remote steal}.
\end{definition}  

Our HBP analysis will proceed as follows.

\begin{enumerate}

\item 
\vhalf
\noindent
{\bf 1. Local Steals.}
We bound the costs of local steals by a straightforward extension of
our previous analysis for BP computations.
For this, we first extend the definition of $f(r)$-data dispersal to 
HBP tasks, and then we bound the cache miss costs for starting task
kernels induced by local steals owned by an HBP task
$\tau$,
to obtain the following lemma
See Appendix~\ref{sec:hbp-local} for the proof.

\begin{lemma}
\label{lem:local-costs-hbp}
Let $\tau$ be a recursive task owning $h$ steals.
Then the cache miss cost for executing $\tau$ is bounded by the
sequential cost plus $(x(\tau)/B + h\cdot B)$, excluding
the costs induced by remote steals.
\end{lemma}

We can also bound  the costs of the
finishing and pseudo task kernels which are
formed by local steals owned by
$\tau$ and which are fully contained in $\tau$.
More precisely,
a finishing or pseudo task kernel is analyzed under local steals only if both the steal at which the task kernel starts and the steal at which it ends are both
local steals. Otherwise, the task kernel is considered to be formed from a remote steal, and its cache miss
overhead is
captured under the analysis for remote steals.
The cost of a task kernel created by a remote steal will be assigned to $\tau$ only
 if the corresponding task kernel is completely contained in $\tau$ and is not contained in any task that is a
 proper descendant of $\tau$.

\item 
\vhalf
\noindent
{\bf 2. Remote Steals.}
Consider a remote steal in an HBP task $\tau$. This will be a local steal in some recursive
task $\tau'$
within $\tau$, but it may create a finishing task kernel and possibly a pseudo task kernel
that contains a portion of  $\tau - \tau'$.
We bound the cost of remote steals in 
Section~\ref{sec:hbp-remote}. Here, we re-assign the cache miss cost of each finishing 
(and pseudo) task kernel $\nu$  to one or two
HBP tasks,
resulting in $O(S)$ different HBP tasks
being charged under this scheme.
For a finishing task kernel, we will see that, in contrast to BP computations,  in the Type $k$ HBP for $k\geq 2$ it can
be more complex than simply a join path, and can itself incur steals.

\item 
\vhalf
\noindent
{\bf 3. Overall Analysis.}
In Section~\ref{sec:anal-summary}, we bound the overall cache miss costs of all local
and remote steals by performing a second level of reassignment of cache miss costs:. 
We define certain {\it special} HBP tasks 
and
we show that we can re-assign the costs that were assigned to $O(S)$  HBP tasks in
Section~\ref{sec:hbp-remote} to at most $4S-1$ special tasks. We show that this 
leads to part $(ii)$ of
Theorem~\ref{thm:cache-miss-overhead-gen}.

\end{enumerate}

\subsubsection{Analyzing Remote Steals}\label{sec:hbp-remote}

The remaining issue is to account for 
remote steals. 
Suppose 
an HBP task $\tau = \tau_1$
incurs a remote steal in one of its recursive constituents, $\mu$. 
Suppose the steal occurs in subtask $\tau_k$, where $\tau_i$ calls $\tau_{i+1}$ recursively, for
$1 \le i < k$, and $\tau_2-\tau_3, \ldots, \tau_{k-1}-\tau_k$ are all steal-free.
This sequence of $\tau_i$ starts at
$\tau_1$, either because it is the root task for the whole computation, 
or because $\tau_1 - \tau_2$ is steal-incurring,
and we will say that $\tau_1$ \emph{adopts} this remote steal and the 
finishing task kernel $\nu$ created by
this steal.

We consider the task kernels for which additional cache misses may occur due to
the presence of this remote steal.
As noted above, any
 starting kernel would be created by a steal in a fork-join tree owned by task $\tau_1$, and hence
would be
 created by a local steal and would be handled by the analysis for local steals.
Thus the only task kernels that may 
incur new costs are pseudo task kernels
and finishing task kernels.
Our analysis will focus on the cost of finishing task kernels.
The analysis of pseudo task kernels is similar.

The finishing task kernel, $\nu$, that emerges from $\tau_k$ will execute portions of 
$\tau_{k-1}-\tau_k, \tau_{k-2}-\tau_{k-1}, \ldots, \tau_2-\tau_3, \tau_1 - \tau_2$, in turn.
In each $\tau_i -\tau_{i+1}$, for $i>1$, the execution of 
$\nu$ starts with the traversal of a path in a join tree
(the join tree complementary to the fork tree from which $\tau_{i+1}$ was forked) followed by
the execution of the remaining constituent tasks in $\tau_i$, if any.
In $\tau_1 - \tau_2$, $\nu$ will traverse a path in the join tree for 
$\mu$
(recall that $\mu$ is the recursive constituent of $\tau$ that contains $\nu$),
and may execute other constituent tasks that follow $\mu$ in $\tau_1$, depending on where the next steal in the computation occurs.
We analyze separately the additional cache miss costs of $\nu$'s access to the join paths,
and $\nu$'s execution of remaining constituent tasks in each $\tau_i$.

In the analysis that follows, we apply a {\it charging} scheme. 
All of the cache miss costs incurred by the finishing task kernel $\nu$ that emerges from a remote steal incurred
by $\tau = \tau_1$ will be 
distributed as charges to  $\tau_1$ and $\tau_2$,
by exploiting the features of cache-compliant HBP algorithms, including the geometric decrease in the extended sizes 
of successive recursive tasks.

\vhalf
{\bf The charging scheme for finishing task kernels that end in
$\tau$}.
First, let us consider
the finishing task kernel $\nu$
described above, and let $\tau$, $\mu$, and the
$\tau_i$ be as described above. We distribute the cache miss costs incurred by $\nu$ as 
follows.

\begin{itemize}
\item[{\bf C1}] The cache miss costs incurred by $\nu$ in $\tau_1 - \tau_2$ are charged to $\tau_1$.

\item[{\bf C2}] All of the remaining cache miss costs (i.e., the costs  for the portions of $\nu$ in $\tau_i-\tau_{i+1}$, for $i>1$) are charged to $\tau_2$.

\end{itemize}

We perform the above distribution for all finishing task kernels that 
start in $\mu$ and end in $\tau$, and we obtain the following two bounds on these charges.
The proofs of these two Lemmas are 
in the Appendix in Section~\ref{app:remote-steals}.

\begin{lemma}\label{cor:C1charge}
Let $\tau$ be an HBP task that incurs remote steals.
Let the $i$-th constituent task of $\tau$ incur $h_i$ steals in its fork tree, and let it incur remote
steals in $c_i$ of its
collection of recursive tasks. Let $c= \sum_i c_i$ and $h = \sum_i h_i$.
Across all steals,
the C1 charge to $\tau$ is bounded by
 $O(x(\tau)/B + B+ h + c)$ to $\tau$.
\end{lemma}

\begin{lemma}\label{cor:C2charge}
Let $\tau_1, \cdots , \tau_k$ be  HBP tasks as defined at the start of Section~\ref{sec:hbp-remote}.
There is a C2 charge of
$O(x(\tau_2)/B + B)$ to 
$\tau_2$  if $\tau_2 - \tau_3$ is steal free
(otherwise $\tau_2 = \tau_k$,
and there is no C2 charge to $\tau_2$).
Across all steals, this is the only C2 charge made to $\tau_2$.
\end{lemma}

\subsubsection{Overall HBP Analysis}
\label{sec:anal-summary}

In Lemmas~\ref{lem:local-costs-hbp}--\ref{cor:C2charge}
 we bounded the additional cache miss
cost of local and remote steals by charging this cost to suitable recursive tasks (or the task that starts the computation).
It remains to determine how many 
different (possibly overlapping)
recursive tasks
 can be charged,
and the amount charged to these tasks, as a function of their extended sizes.
Once we have obtained good bounds for these, 
we will readily obtain the desired bound in part $(ii)$ of Theorem~\ref{thm:cache-miss-overhead-gen}.

There are at most $S$ recursive tasks that incur a
local steal, i.e., a steal in a fork tree they own.
We call these \emph{Type 1 special tasks}. 
It is convenient to make the root task, the task which starts the computation, Type 1 also.
There are at most a further $S-1$ tasks which have steals in two or more of the recursive subtasks they
call, while having no steals in their fork trees.
We call these \emph{Type 2 special tasks}. 
The Type 1 and Type 2 tasks correspond to the $\tau_1$'s in the 
analysis in Section~\ref{sec:hbp-remote}; also, local steals occur only in Type 1 tasks.
The remaining charged tasks, corresponding to the $\tau_2$ when $k > 2$, 
must all be a child of a Type 1 or Type 2 task and further must have a descendant of Type 1 or 2 
(corresponding to the $\tau_k$).
We call these \emph{Type 3 special tasks}. 
Thus there are at most 
$2S-1$ of these,
yielding a total of $4S-1$ special tasks.  
Note that only the special tasks have been charged.

We now perform one more round of redistribution of costs in order to remove the $c$ term
from Lemma~\ref{cor:C1charge}. 
This will result in a charge to each special task that depends only
on its extended size and the number of local steals.

\begin{lemma}
\label{lem:redistr}
The charges to the special tasks for all steals
can be redistributed so that a Type 1 task $\tau_1$ that owns
$h$ steals receives a charge of $O(x(\tau_1)/B + B + h\cdot B)$,
a Type 2 task $\tau'_1$ receives a charge of $O(x(\tau'_1)/B + B)$
and a Type 3 task $\tau_2$ receives a charge of  
$O(x(\tau_2)/B + B)$.
\end{lemma}

Summing over all charged tasks, and noting that the number of special tasks is at most
$4S-1$ (as shown above) yields a total charge of
$O(\sum_{1 \leq i \leq O(S)} x(\tau_i) + S\cdot B),$
where the $\tau_i$ are distinct recursive or BP tasks.

Together with the analysis of the costs due to pseudo task kernels,
this proves the second claim in Theorem~\ref{thm:cache-miss-overhead-gen}.

\vhalf
\noindent
{\bf Acknowledgement.} We thank Charles Leiserson for extensive discussions and helpful comments.
We also thank Simon Peters and Emmett Witchel for their comments on schedulers used in practice,
and Phil Gibbons for drawing our attention to the potentially larger constant factor in the $O(Q)$ term
in our 
bound for general schedulers compared to 
that for
work-stealing
schedulers in prior work~\cite{Blumofe96b,ABB02}.

%% file: alg-anal-app.tex
\section{Analysis of Specific Algorithms}
\label{sec:alg-anal-app}
We apply Theorem~\ref{thm:cache-miss-overhead-gen} to several well-known algorithms,
to obtain the results in Table~\ref{table1}. The GEP and LCS algorithms are presented in~\cite{CR07,CLR10}, while the others
 are described in~\cite{CR12}
(where their false sharing costs
were analyzed).

\vhalf
\noindent
{\bf Scan, Prefix Sums, Matrix Transpose (MT).} These algorithms are 
Type 1
HBP, so  
Theorem~\ref{thm:cache-miss-overhead-gen} directly gives the bound for MT. The bound for Scan and Prefix Sums
follows from a more careful
analysis given in Section~\ref{sec:bp}.

\vhalf
\noindent
{\bf $\log^2 n$-MM.} This is a
Type 2
HBP that has one recursive constituent that makes 8 recursive calls to$n/2\times n/2$ matrice,
 and a BP task that adds up the outputs of the recursive calls in pairs. Applying 
Theorem~\ref{thm:cache-miss-overhead-gen}, we see that  we will have the $O(l)$ largest HBP tasks by including all recursive tasks up to $j=(1/3) \log l$ levels of recursion. The sum of the sizes of these tasks is
$O((n^2/4^j) \cdot 8^{(1/3) \log l})
 = O(n^2 \cdot l^{1/3})$. Since $l=O(S)$,  we obtain
  the overall cache miss cost with $S$ steals as
 $O(Q + (n^2/B) \cdot S^{1/3} + S \cdot B)$.
 
 \vhalf
 \noindent
 {\bf Depth-n-MM.} 
 This algorithm from~\cite{FLPR99} is a Type 2 HBP that has two recursive constituents, each of
 which makes 4 calls to $n/2\times n/2$ matrices. Since there are
 $8^j$ instances of size $n/4^j$, it has the same properties
as $\log^2 n$-MM for applying Theorem~\ref{thm:cache-miss-overhead-gen}, and its cache miss bound under a general scheduler remains the same.

 \vhalf
 \noindent
 {\bf GEP.}  Gaussian Elimination Paradigm (GEP) and its in-place variant I-GEP~\cite{CR07} are Type 4 HBP.
They solve several important problems including
Gaussian elimination without pivoting and graph all-pairs
shortest paths, and a degenerate form of I-GEP reduces to depth-n-MM.
Although considerably more complex than Depth-n-MM, I-GEP creates 8  $n/2\times n/2$ recursive subproblems and hence has the same caching bound as Depth-n-MM
with $S$ steals. 
 
\vhalf
\noindent
{\bf Strassen's Matrix Multiplication.}
This algorithm is structurally the same as $\log^2 n$-MM, except that is creates 7, not 8, recursive calls with
$n/2$ rows and columns.
It has  $Q= O(n^{\lambda}/(B \cdot M^{\gamma})$, where 
$\lambda=\log_2 7$ and $\gamma = (\lambda/2)-1$~\cite{FLPR99,CR11}.
Applying the same analysis as for $\log^2 n$-MM, we obtain 
$C(S) = O(Q + (n^2/B) \cdot S^{1- (2/\lambda)} + S \cdot B)$.

\vhalf
\noindent
{\bf FFT, SPMS Sort.} The algorithms for both FFT~\cite{FLPR99} and 
SPMS sort~\cite{CR10} have the same structure, being Ttpe 2 HBP algorithms that 
recursively call two collections of $O(\sqrt n)$ parallel tasks of size $O(\sqrt n)$, together
with a constant number of calls to BP computations.

It remains to bound the term
$ \sum_{i=1}^{O(S)} |\nu_i|/B$.
The total size of tasks of size $r$ or larger is $O(n \log_r n)$, and there are
$\Theta(\frac{n}{r} \log_r n)$ such tasks.
Choosing $r$ so that $S = \Theta(\frac{n}{r} \log_r n)$,
implies that $r\log r = \Theta( n\log n/S)$, so
$\log r = \Theta(\log( [n \log n/S])$.
Thus  $ \max_{\cal C} \sum_{\nu_i\in {\cal C}}\frac{|\nu_i|}{B}
= O(\frac{n}{B} \log_r n) = O(\frac{n}{B} \frac{\log n}{\log [(n\log n)/S]})$.

The analysis for SPMS is very similar, except that we need to handle two BP computations
with somewhat irregular access patterns. 

\vhalf
\noindent
{\bf List and Graph Algorithms}
The list ranking algorithm (LR) described in ~\cite{CR12}
has $Q = O( \frac nB \log_M n)$ sequential cache complexity.
It
begins with $\log \log n$ phases which reduce the list length by an $O(\log n)$ factor.
In the $i$th phase there are $k \ge 2$ sorts of size $O(n/2^i)$ followed by
a sequence of $\log^{(k)} n$ sorts of combined size $O(n/2^i)$, where $k\ge 2$ is a
constant. So the number of these sorts
is $O(\log \log n \cdot \log^{(k)} n) = O((\log \log n)^2)$.
The remaining phase, on a list of size $O(n/\log n)$, uses the standard
parallel algorithm.
Each step of this
algorithm requires a sort of size $O(n/\log n)$.
Thus LR comprises a sequence of
$O(\log n)$ sorts of combined size $O(n)$.

If the $i$th sorting problem has size $n_i$ and incurs $S_i$ steals,
using the SPMS cache miss bound, 
the bound on cache misses is
$O(Q + S\cdot B + \sum_i \frac{n_i}{B} \frac {\log n_i} {\log [((n_i\log n_i)/S_i]})
= O(Q + S\cdot B + \sum_i \frac{S_i}{B}\frac{(n_i \log n_i)/S_i} {\log [((n_i\log n_i)/S_i]})$.
Since $x/\log x$ is a concave function, this sum is maximized
when $n_i\log n_i/S_i = n'/S = O(n/S)$
where $S= \sum_i S_i$ and $n' = \sum_i n_i$.
This yields a bound of
$O(Q + S\cdot B + \frac{n}{B} \frac {\log n} {\log [((n\log n)/S]})$ cache misses.

Likewise, the most expensive part of the connected components algorithm is $O(\log n)$ iterations
of LR~\cite{CR12}, which yields a bound of
$O(Q + S\cdot B + \frac{n+m}{B} \frac {\log^2 n} {\log [((n+m)\log n)/S])}$ cache misses,
where $n$ in the number of vertices and $m$ the number of edges in the graph.

\vhalf
\noindent
{\bf Longest Common Subsequence (LCS)}.
LCS~\cite{CR08,CLR10} is a Type 3 HBP that has a constituent that is a
 Type 2 HBP that finds only  the length of an LCS, and one recursive constituent with three 
 parallel calls of size $n/2$. There are $O(4^j)$ tasks of size $n/2^j$. Applying
 Theorem~\ref{thm:cache-miss-overhead-gen} we obtain the cache miss bound as
 $O(Q + (n/B) \cdot \sqrt S + S \cdot B)$.

\vone
\noindent
{\bf Two Observations on our Bounds.}

\vhalf
1. {\bf Improved Bound for FFT and SPMS.}
For FFT and SPMS, our refined bound is strictly better than the 
$O(Q + S \cdot M/B)$
 bound since
our overhead remains $O(Q)$ when $M ^{\epsilon} = O((n \log n)/S$ for any constant $\epsilon>0$, 
while the simple bound needs $M \log M = O((n \log n)/S)$ for $Q$ to dominate $S \cdot M/B$.

\vhalf
2. Although Depth-n-MM and I-GEP have the same cache miss bound as a function of the number of
steals, under widely used schedulers such as randomized work stealing (RWS), the expected cache miss cost for
I-GEP will be larger since I-GEP's critical path length $(T_{\infty}$) is larger ($n \log^2 n$ versus $n$ for Depth-n-MM),
and the expected number of steals depends on $T_{\infty}$ for RWS. The same distinction applies to List Ranking (LR) versus
SPMS.

%% file: data-examples-app.tex
\section{HBP Analysis}

\subsection{Data Dispersal Function for Some Algorithms}\label{sec:f-r-examples}

We illustrate the definition of $f(r)$ (see Section~\ref{sec:data}) with some fairly detailed descriptions of 
several standard cache efficient  BP computations.

\vhalf
1. {\bf Scan and Sum.} 
A scan of an array $A[1..n]$ stored in contiguous locations in $\lceil n/B\rceil$ blocks can be performed 
as a BP computation by
recursively forking calls to Scan on the left and right halves of the array. 
The base case at size 1 reads the element in that size 1 subarray.
Here, the elements are accessed
only at the leaves of the $n$-leaf BP computation, 
and there is no access to the input data at the internal fork and join nodes. 
Thus an inorder traversal of the fork tree corresponds to a left to right ordering of the leaves, 
and any extended task of size $r$ will
access at most $\lfloor r/B\rfloor +2$ blocks, where the +2 accounts for 
possible partial blocks
 at the beginning and end of the sequence of elements
in $A$ accessed by the extended task. 
Thus, this BP computation has $f(r) = O(1)$.

A similar computation outputs the sum of the elements in array $A[1..n]$. 
Here, a join node computes the sum of the elements at the leaves of its subtree by adding the sum at its two children,
with the addition being performed by the children adding their sums to an initially zero value at the parent.
The overall sum is the value computed at the root of the join tree. In fact, this computation can output
an array $S[1~..~2n-1]$, where $S[i]$ contains the sum of the elements in the subtree rooted at the
node with inorder number $i$.

\vhalf
2. {\bf Prefix Sums.} 
We compute prefix sums of an input array $A[1..n]$ using the following 3-phase algorithm. This has
$f(r)= O(1)$ as described above.

{\it Phase 1:} Compute the sum array $S[1..2n-1]$ as in 1 above. This has $f(r) = O(1)$.

{\it Phase 2:} Compute the array $LS[1..2n-1]$, where $LS[i]$ computes the sum of the values at 
leaves that lie to the left of the subtree rooted at the node with inorder number $i$.

{\it Phase 3:}  Compute the output array $PS[1..n]$ by computing $PS[x]$ at each leaf $x$ 
as $LS[x] + A[x]$. This accesses only data
in node $x$'s segment on the execution stack plus the 
item in $A$ associated with $x$;
trivially it has $f(r)=O(1)$.

\vhalf
Consider Phase 2. A straightforward method for Phase 2 is as follows: For the root $r$ set 
$LS[r] \leftarrow 0$.
For an internal node $y$ that is a left child of its parent $x$, $LS[y] \leftarrow  LS[x]$, and for the right
child $z$ of $x$, $L[z] \leftarrow LS[x] + S[y]$, since $y$ is the left sibling of $y$. However, this computation
does not satisfy the local constraint,
 and in fact, $f(r) = \log r$ (to be precise $f(r) = \log (r/B)$) since we
could have an extended task $\mu$ of size $r$ where up to $\log r$  of the left siblings for nodes
within $\mu$ may lie in blocks not contained within $\mu$.

Here is an alternate implementation of Phase 2 that has $f(r) = O(1)$.
  First, in Phase 1, also
compute the array $SC[1..n-1]$, where, for the $i$-th internal node in the fork tree 
 (in inorder), $SC[i]$ contains the sum value computed at  the
left child of $i$. The value
 for this second array can
 be copied over to the parent node on the execution stack when the sum value is computed at the
 join node for the  left child, 
but the parent node will be the one to store this value in $SC[i]$. With this implementation this computation 
satisfies the local constraint,
 hence $f(r) = O(1)$. Then in Phase 2 
 if node $x$ has left child $y$ and right child $z$ in the fork tree, then
$LS[y] \leftarrow LS[x]$ and $LS[z] \leftarrow LS[x] + SC[x]$. 
As with the computation of $SC$, the values $LS[x]$ and $SC[x]$  are copied over to the execution stack
by node $x$, 
and the computation at a node $w$ stores the value in shared array location
$LS[w]$ (where $w$ is $y$ or $z$).
This is a BP computation 
that satisfies the local constraint,
 and has $f(r) = O(1)$.

In the above implementation, a node  $v$ 
in the join tree
accesses data located both at $v$ and the parent of $v$. 
This access pattern, which 
is said to satisfy the {\it local constraint}~\cite{CR12},
can be seen to give $f(r)= O(1)$ if 
the algorithm stores the data at the
parent of $v$ that is accessed by $v$ on the execution stack (as is the case above).
This gives rise to a contiguous data access pattern since the data at $v$ and at the parent of $v$ are contiguous on the execution stack.  If this data was
 stored in inorder in a shared array, the data for $v$ and $v$'s parent would not be contiguous in general.

\vhalf
3. {\bf Matrix Transpose.} To transpose an $n \times m$ matrix stored in row major order, 
we use a BP computation that forks two submatrices by splitting along the
longer dimension (i.e, into two subproblems with the first and last $n/2$ rows if $n \geq m$, 
and otherwise splitting along the columns)~\cite{FLPR99}.
The base case at size 1 copies the element at position $(i,j)$ in the input matrix into position $(j,i)$ in the output matrix.
For a highly skewed rectangular matrix where $n>2m$ or $m>2n$, 
this process repeatedly halves along the larger dimension until the ratio of
the number of rows to number of columns is between 1/2 and 2, 
and thereafter it alternates between halving rows and halving columns.

As with Scan, all data accesses occur at the leaves of the BP computation, 
so the data accesses are in left to right order along the leaves. 
But here, in contrast to Scan, the data layout is in row major
order for the input $n \times m$  matrix. We analyze the case when $n$ and $m$ are
within a factor of 2 of each other. Consider 
an extended task $\tau$ of size $r$, 
and let $z$ be an internal node (possibly the root)
of the fork tree whose subtree contains all of the leaves in $\tau$. 
Let $r_1$ of $\tau$'s
 leaves be in the left subtree of $z$, and
$r_2 = r- r_1$ in the right subtree. 
Let us consider the $r_1$ leaves accessed in
 the left subtree, and let
$x$ be the lowest internal node in the
BP fork tree whose subtree contains all of these $r_1$ nodes. 
Then, since these $r_1$ nodes extend to the rightmost leaf in this subtree,
$r_1$ is larger than half the total number of leaves $n_1$ in this subtree. 
Let the submatrix represented by this subproblem have dimension $p \times q$.
Then the number of blocks accessed is $p \cdot q + p$ since the matrix is stored in row major order.
If $q\geq p$ then $f(r_1) = O(\sqrt r_1)$.
Otherwise, we consider the subproblem rooted at 
 $y$, the parent of $x$. 
This subproblem has size $2 \cdot p \cdot q$, and since $p>q$, it must
have dimensions $p \times 2q$ with $2q > p$. 
Thus, $f(r_1)$ remains $O(\sqrt r_1)$
 in the case $p<q$ as well. 
A similar analysis holds for the
$r_2$ nodes in
 the right subtree of the root problem. 
It is also readily seen that $f(r) = O(\sqrt r)$ for the initial phase of halving a
highly skewed rectangular matrix. Thus, Matrix Transpose has $f(r) = O(\sqrt r)$.

\vone
The algorithms we analyze 
(except for two procedures in SPMS sorting)
are all 
$O(\sqrt r)$-dispersed, and some are $O(1)$-dispersed (for which we give an
improved bound).
As a result they satisfy our assumption in Section~\ref{sec:results} that a task that 
accesses $r$ words
will access $O((r/B) + \sqrt r)$ blocks. We present this more general analysis here since it allows one 
to fully analyze the SPMS algorithm and other algorithms with complex data access patterns.

%% file: local-steal-anal-app.tex
\subsection{Analysis of Local Steals}\label{sec:hbp-local}

We extend the definition of $f(r)$-dispersion
 to HBP tasks.
Consider a recursive constituent $\mu$ in an HBP task $\tau$.
Recall that a recursive constituent comprises a collection of recursive tasks that can be executed in parallel,
together with the fork tree forking these tasks and the complementary join tree.
The variables declared by $\tau$ and the algorithm's input and output variables are viewed
as the shared data structures (or global variables)
for the constituents within $\tau$, in contrast to the synchronization
variables declared by its fork nodes (variables needed to manage the forking and joining)
and the variables declared by the individual constituents
and the recursive tasks they contain;
these variables are not shared, and are on the execution stack.

\begin{definition}
\label{def:hbp-cache-fr}
{\bf (Data dispersal function $f(r)$ for HBP)}
Let $\mu$ be a recursive constituent in an HBP task $\tau$, and let 
$\sigma_1, \sigma_2, \ldots, \sigma_k$ be the 
sequence of parallel 
recursive tasks forked by $\mu$'s fork tree.
Let $\sigma_i, \ldots, \sigma_j$ be an arbitrary consecutive subsequence of $\sigma_1, \ldots, \sigma_k$,
and let $r$ be the size of the computation represented
by these nodes.
If this computation accesses $r/b + f(r)$ blocks 
in the shared data structures,
it is said to be $f(r)$-dispersed.
$\mu$ is $f(r)$-dispersed
if this property holds for every consecutive subsequence
of its recursive tasks,
and $\tau$ is $f(r)$-dispersed
if this property holds for every recursive constituent task,
and every BP task.
Finally, an HBP computation  is $f(r)$-dispersed
if this property holds for every BP and recursive HBP task in its computation.
\end{definition}

For example, in the FFT algorithm, there is a recursive constituent that makes $\sqrt n$ calls to an algorithm for MT
(matrix transposition), with each subproblem having size $\sqrt n$, and hence representing an
$n^{1/4} \times n^{1/4}$ matrix. The algorithm assumes that the input representation stores the overall matrix in row-major.
Then each submatrix has $f(r) = \sqrt r$, and the algorithm has the submatrices corresponding to a consecutive
 subsequence of $k$ calls  correspond to a $\sqrt k \times \sqrt k$ packing of these $k$ matrices in the input matrix. Thus $f(r)$ continues to be $O(\sqrt r)$ for this HBP computation
for FFT. Another class of examples is provided by algorithms
where each recursive constituent has a constant size fork tree (as in the matrix multiplication algorithms);
these algorithms will inherit the $f(r)$ due to
 a single recursive call.

We will assume that the HBP computation is cache-compliant (Definition~\ref{def:good-hbp}) 
and, as before, that $f(r) = O(\sqrt r)$.
With this assumption, the cache miss cost of executing the subsequence 
$\sigma_i, \ldots, \sigma_j$ in the above definition is $O(\sum_{l=i}^j x(\sigma_l) + f(r))$.

The analysis for local steals from a
constituent BP task $\mu$
 in an HBP task $\tau$ is exactly the same
as our earlier analysis for BP 
computations.
Its cost, if it incurs $S_{\mu}$ steals, is $O(|\mu|/B +S\cdot B)$ or $O(|\mu|/B + S)$,
according as $f(r) = O(\sqrt r)$ or $f(r) = O(1)$,
and summed over all the constituent BP tasks, this totals $O(x(\tau) + S\cdot B)$
or $O(x(\tau) + S)$, respectively,
where $S$ is the total number of steals incurred by these BP constituents.

 So we only consider $\tau$'s constituent recursive tasks  here.
In particular, let $\mu$ be a single recursive constituent task and let
$\sigma_1, \sigma_2, \ldots, \sigma_k$ be the 
sequence of parallel 
recursive tasks forked by $\mu$'s fork tree.
Suppose that $\mu$'s fork tree incurs $h$ steals.
We can bound the cache miss excess as in
 the BP analysis. The portions of the
finishing task kernels within $\mu$ only execute paths in $\mu$'s join tree, and their analysis
is exactly as in the BP case. There are at most $3h+1$ starting and pseudo task kernels 
induced by the $h$ steals
(by Lemma~\ref{lem:task-kernel-number}),
 and the cache miss cost for these task kernels totals
$O(x(\mu) +  \sum_{i=1}^{3h+1}\sqrt{r_i})
= O(x(\tau) + \sum_{i=1}^{3h+1}\sqrt{r_i})$,
where the $r_i$ are the sizes of the task kernels created by the $h$ steals.
Since  each task kernel is formed from
 one of more of the complete $\sigma_l$,
 $\sum_{i=1}^{3h+1} r_i = \sum_{l=1}^k |\sigma_l| =O(x(\tau))$,
explaining the definition of $x(\tau)$
(see Definition~\ref{def:extended-size}).
As in the case of
BP computations,
$ \sum_{i=1}^{3h+1}\sqrt{r_i} = O(x(\tau)/B + h\cdot B)$.

The costs for accessing the execution stacks are bounded just as for BP tasks.
The one additional feature that we need to address is that the final finishing task kernel in a
constituent task that incurs steals may incur additional cache misses 
in executing the remainder of $\tau$, but 
since $f(r) = O(\sqrt r)$,
no more than 
$O(x(\tau)/B + \sqrt{x(\tau)})$.
Even if several of $\tau$'s constituents incur steals in their fork trees, but nowhere else,
as there are only $O(1)$ constituents, the total charge is still 
$O(x(\tau)/B + \sqrt{x(\tau)})$
for the final finishing task kernel in each constituent that incurs steals, except the last. The final
finishing task kernel $\mu$ (for the last local steal in $\tau$) may not terminate within $\tau$. In that case,
we do not bound the cache miss cost for this task kernel here under local steals. The cost for $\mu$ 
will be bounded under remote steals.

In sum, for $\tau$'s  local steals (i.e., for steals owned by $\tau$) we have:

\vhalf
\noindent
{\sc Lemma~\ref{lem:local-costs-hbp}}~
\emph{Let $\tau$ be a recursive task owning $h$ steals.
Then the cache miss cost for executing $\tau$ is bounded by the
sequential cost plus $(x(\tau)/B + h\cdot B)$, excluding
the costs induced by remote steals.}
\begin{proof}
The discussion above has shown a bound of \\
$O(x(\tau)/B + h\cdot B + \sqrt{x(\tau)})$.
We now observe that if $x(\tau) \ge B^2$ then 
$\sqrt{x(\tau)} \le x(\tau)/B$ and otherwise $\sqrt{x(\tau)} \le B$.
\end{proof}

This analysis still applies if $\tau$ incurs remote steals,
except that it does not cover the costs induced by those remote steals.
We address this next.
Also, note that any starting kernel starts at the right child of a node in a fork tree, and hence its cost is charged as
 a local steal cost to
 the HBP task that owns that starting kernel. Hence, the analysis for remote steals in the next section needs to consider only finishing and pseudo task kernels.

%% file: remote-steal-anal-app.tex
\subsection{Analysis of Remote Steals}\label{app:remote-steals}

We now establish Lemmas~\ref{cor:C1charge} and \ref{cor:C2charge}.
The analysis is structured as follows. In Lemma~\ref{lem:anal-hbp-join-paths}
we bound the C2 charge to $\tau_2$ for 
accesses  to the join paths in $\nu$, and in 
Lemma~\ref{lem:const-tau-two-charge} we bound the 
C2 cost of accesses by $\nu$
outside of these join paths; in contrast to 
the analysis for a finishing task kernel in
a BP computation, these latter accesses can be
substantial in an HBP computation, and for each $i$, they will include the execution of
all constituent tasks that follow 
the first constituent that overlaps
 the portion of $\nu$ in $\tau_i$. In 
Lemma~\ref{lem:join-tau-one-charge}  we bound the 
C1 charge to $\tau_1$ for 
accesses to the join path in a recursive constituent $\mu$ for
all finishing task kernels
adopted by $\tau_1$ that are induced by steals in $\mu$ (since there are only $O(1)$
recursive constituents this also gives a bound on the total C1 charge to $\tau_1$ for accesses
to the join paths).
Finally, in Lemma~\ref{lem:const-tau-one-charge} we bound the 
C1 charge to $\tau_1$ for executing all
finishing task kernels 
adopted by $\tau_1$,
outside of the accesses to the join trees. 
Lemmas~\ref{cor:C1charge} and \ref{cor:C2charge} 
then simply
summarize the overall charges applied by this
charging scheme to $\tau_1$ and $\tau_2$ respectively.

\vhalf
\noindent
{\bf 
Charges for $\nu$'s accesses to join paths.}

\begin{lemma}
\label{lem:anal-hbp-join-paths}
The C2 charge to $\tau_2$ by finishing task kernel $\nu$ for accesses to the join paths is
  $O(x(\tau_2)/B + \log B)$.
\end{lemma}
\begin{proof}
The length of the join path in $\tau_i - \tau_{i+1}$ is
$O(\log x(\tau_i))$, and as the segments for its nodes
are consecutive on the execution stack, its execution
induces $O(\ceil{(\log x(\tau_i)/B})$ cache misses.
By assumption, the $x(\tau_i)$ are geometrically increasing,
so for $i \ge 2$, $\log x(\tau_i) = O(\log x(\tau_2))$,
and $k$, the number of levels of recursion is $ O(\log x(\tau_2))$.
Hence the first $k-1$ paths (i.e.\ excluding the path in $\tau_1 - \tau_2$), incur $O(\log^2 x(\tau_2)/B + \log x(\tau_2))$
cache misses.
This will be charged to $\tau_2$.
This charge is at most $O(x(\tau_2)/B + \log B)$; to see this note that if $x(\tau_2) \le B^2$,
then $\log x(\tau_2) = O(\log B)$ and otherwise $\log x(\tau_2) = O(\log (x(\tau_2)/B) )= O(x(\tau_2)/B )$.
\end{proof}

We now bound the charge to $\tau_1$ for accesses to the join paths in the join tree for a single recursive
constituent $\mu$ by all finishing task kernels that traverse this join tree and end in $\tau_1$.

\begin{lemma}
\label{lem:join-tau-one-charge}
Let $\mu$ be a recursive constituent task of task $\tau$
 and
suppose its fork tree incurred $h$ steals.
Further suppose $\mu$ incurs 
remote steals in $c$ of its collection of recursive tasks.
Then the C1 charge to $\tau_1$ for executing the join paths in $\mu$'s join tree
is bounded by $O(x(\tau_1)/B + h +c)$ cache misses.
\end{lemma}
\begin{proof}
We use the same argument as in the analysis in Lemma~\ref{lem:bp}.
Here $f(r) = 0 = O(1)$ for there are no accesses from the join tree to the shared data structures.
The one change is that each of $\mu$'s steal-incurring recursive tasks adds one more
path to the join tree for a total of $c$ extra paths.
This creates $c$ additional cache misses due to the resulting discontinuities at the end of each
path thereby raising the total cost by $c$.
\end{proof}

The following corollary generalizes the above lemma to include steals in all constituents of $\tau$,
and it readily holds since there are only $O(1)$ constituents in $\tau$.

\begin{corollary}\label{cor:join-tau-one-charge}
Let $\tau$ be an HBP task that incurs remote steals.
Let the $i$-th constituent task of $\tau$ incur $h_i$ steals in its fork tree and  remote
steals in $c_i$ of its
collection of recursive tasks. Let $c= \sum_i c_i$ and $h = \sum_i h_i$.
Across all steals,
the C1 charge to $\tau$ for executing the join paths in its constituents is bounded by
 $O(x(\tau)/B +  h + c)$
\end{corollary}

\vhalf
\noindent
{\bf 
Charges for $\nu$'s accesses to the $\tau_i - \tau_{i-1}$,
outside of the join paths}.
%
\begin{lemma}
\label{lem:const-tau-two-charge}
The C2
 cost for executing the sequence of recursive constituents in $\tau_{k-1} - \tau_k, \ldots,
\tau_2 - \tau_3$ is bounded by $O(x(\tau_2)/B + \sqrt{x(\tau_2)} +\log B)
= O(x(\tau_2)/B + B)$ and 
is charged to $\tau_2$.
\end{lemma}
\begin{proof}
The result follows from the geometric
 decrease in the sizes of the $\tau_i$ as $i$
increases.
Since $f(r) = O(\sqrt r)$, 
the cost of executing the recursive constituents of $\tau_i - \tau_{i+1}$ is 
$O(x(\tau_i)/B + \sqrt{x(\tau_i)})$ cache misses.
Consequently the cost for the sequence of $k-1$ recursive constituents is
$x(\tau_{k-1})/B + O(\sqrt{x(\tau_{k-1}}) + \ldots + x(\tau_2)/B  + O(\sqrt{x(\tau_2)}) = O(x(\tau_2)/B + \sqrt{x(\tau_2)} + k) = O(x(\tau_2)/B + \sqrt{x(\tau_2)}+ \log x(\tau_2))
=  O(x(\tau_2)/B  +B)$,
as the $x_i$ grow geometrically with decreasing $i$.
\end{proof}

\begin{lemma}
\label{lem:const-tau-one-charge}
The C1 charge to $\tau_1$ by finishing task kernels 
induced by steals in the fork trees for
 its recursive constituents
is at most $O(x(\tau_1)/B + \sqrt{x(\tau_1)}) = O(x(\tau_1)/B + B)$.
\end{lemma}
\begin{proof}
Consider the finishing task kernel $\nu$ defined earlier.
The cost for $\nu$'s computation of these recursive constituents is at most $O(x(\tau_1)/B + \sqrt{x(\tau_1)})$ cache misses.
As at most one finishing task can start each of $\tau_1$'s 
$O(1)$ constituent
tasks, the total charges of this type sum to $O(x(\tau_1)/B + \sqrt{x(\tau_1)})$.
\end{proof}

Lemma~\ref{cor:C1charge} (repeated below) follows from 
Corollary~\ref{cor:join-tau-one-charge}  and 
Lemma~\ref{lem:const-tau-one-charge}, and
Lemma~\ref{cor:C2charge} (also repeated below)
follows from Lemmas~\ref{lem:anal-hbp-join-paths} and~\ref{lem:const-tau-two-charge}.

\vhalf
\noindent
{\sc Lemma~\ref{cor:C1charge}}.
\emph{Let $\tau$ be an HBP task that incurs remote steals.
Let the $i$-th constituent task of $\tau$ incur $h_i$ steals in its fork tree, and let it incur remote
steals in $c_i$ of its
collection of recursive tasks. Let $c= \sum_i c_i$ and $h = \sum_i h_i$.
Across all steals,
the C1 charge to $\tau$ is bounded by
 $O(x(\tau)/B + B+ h + c)$ to $\tau$.}

\vhalf
\noindent
{\sc Lemma~\ref{cor:C2charge}}.
\emph{Let $\tau_1, \cdots , \tau_k$ be  HBP tasks as defined at the start of Section~\ref{sec:hbp-remote}.
There is a C2 charge of
$O(x(\tau_2)/B + B)$ to 
$\tau_2$  if $\tau_2 - \tau_3$ is steal free
(otherwise $\tau_2 = \tau_k$,
and there is no C2 charge to $\tau_2$).
Across all steals, this is the only C2 charge made to $\tau_2$.
}

%% file: proof-lem-redistr-app.tex
\subsection{Proof of Lemma~\ref{lem:redistr}}

\vhalf
\noindent
{\sc Lemma~\ref{lem:redistr}}
\emph{The charges to the special tasks for all steals
can be redistributed so that a Type 1 task $\tau_1$ that owns
$h$ steals receives a charge of $O(x(\tau_1)/B + B + h\cdot B)$,
a Type 2 task $\tau'_1$ receives a charge of $O(x(\tau'_1)/B + B)$
and a Type 3 task $\tau_2$ receives a charge of  
$O(x(\tau_2)/B + B)$.
}

\begin{proof}
Each Type 1 or Type 2 task redistributes  its $O(c)$ term, giving $\Theta(1)$
units to each Type 1 or Type 2 descendant at the bottom of a path of recursive calls defined above 
(namely the $\tau_k)$. 
Each such $\tau_k$ receives only one $\Theta(1)$ charge, so the overall charge to a special
task $\tau$ incurring $h$ steals in its fork trees is $O(|\tau|/B + B + h)$ for all remote steal
costs and for the finishing task kernels for the local steals. Finally by
Lemma~\ref{lem:local-costs-hbp}, the total cost for the local steals for a Type 1 special task
is $O(x(\tau_1)/B + h \cdot B)$, and this establishes the lemma.
\end{proof}

%% file: pseudo-kernel-app.tex
\subsection{Analysis of Pseudo Task Kernels} 

The above analysis bounded the caching overhead for finishing task kernels formed by remote steals.
The costs due to pseudo kernels $\nu$ 
formed by remote steals
can be bounded similarly.
Again, there may be a recursive sequence $(\tau_1, \tau_2, \ldots, \tau_k)$,
where $\tau_i$ calls $\tau_{i+1}$, for $1 \le i < k$, and $\nu$ is induced by a steal in $\tau_k$
but starts in $\tau_1 - \tau_2$.
Then $\nu$ will incur costs for executing a portion of $\tau_1 - \tau_2$, and for
executing portions of $\tau_i - \tau_{i+1}$, for $2 \le i < k$, and for executing a portion of
$\tau_k$. The latter costs total $O(x(\tau_2)/B +B)$ and are charged to $\tau_2$.
The former costs are charged to $\tau_1$, and summed over all $h$ pseudo tasks making
such charges to $\tau_1$, total $O(x(\tau_1) + h\cdot B)$.
It is not hard to see these charges are no larger than those induced by finishing tasks,
thus the bound given in Lemma~\ref{lem:redistr}
continues to apply.